\documentstyle[multicol,prb,aps]{revtex}

\begin{document}
\draft
\title{Diffusion theory of spin injection through resistive contacts}
\author{Emmanuel I. Rashba\cite{Rashba*}}    
\address{Department of Physics, The State University of New York at Buffalo, Buffalo, NY 14260}  
\date{May 1, 2002}
\maketitle
\begin{abstract} 
Insertion of a resistive contact between a ferromagnetic metal and a semiconductor microstructure is of critical importance for achieving efficient spin injection into a semiconductor. However, the equations of the diffusion theory are rather cumbersome for the junctions including such contacts. A technique based on deriving a system of self-consistent equations for the coefficients of spin injection, $\gamma$, through different contacts are developed. These equations are concise when written in the proper notations. Moreover, the resistance of a two-contact junction can be expressed in terms of $\gamma$'s of both contacts. This equation makes calculating the spin valve effect straightforward, allows to find an explicit expression for the junction resistance and to prove that its nonequilibrium part is positive. Relation of these parameters to different phenomena like spin-e.m.f. and the contact transients is established. Comparative effect of the Coulomb screening on different parameters is clarified. It is also shown that the spin non-conservation in a contact can have a dramatic effect on the non-equilibrium resistance of the junction.           
\end{abstract}


\begin{center}
{\it To appear in Euro. Phys. Journal B}
\end{center}

\begin{multicols}{2}
\narrowtext
\section{Introduction}               
\label{sec:intro}

Efficient spin injection from a ferromagnetic (F) conductor into a non-magnetic (normal, N) conductor is one of the central problems of spintronics.\cite{Wolf,DSFHZ} Achieving spin injection into semiconductor microstructures is a prerequisite for building up a spin transistor proposed by Datta and Das,\cite{DD90} that is a spin-interference device based on the spin precession controlled by a gate voltage {\it via} spin-orbit interaction.\cite{R60} This problem is closely related to the general problem of the passage of an electric current through a F-N-interface that is also of primary importance for the giant magnetoresistance (GMR)\cite{GMR} and the devices based on it. In the heart of the problem is the concept of the coupling between the charge and spin currents put forward by Aronov\cite{A76} and developed by Johnson and Silsbee in terms of the thermodynamics of irreversible processes.\cite{JS8788} These authors have also demonstrated the spin injection through the interface of F- and N-metals that has been detected by the open-circuit voltage at a second F-electrode (spin-e.m.f., electromotive force).\cite{JS85} Before long, a diffusion theory of the spin-injection through a F-N-interface has been also developed\cite{vS87} and successfully applied to the current-perpendicular-to-the-plane (CPP) geometry of GMR.\cite{VF93} 

The success in injecting spins from metallic ferromagnets into paramagnetic metals through ``perfect" (Ohmic, non-resistive) contacts was followed by less encouraging results when applied to the spin injection into semiconductors.\cite{AP76} Spin injection at the level of only about 1\% has been reported,\cite{Ham99,Gard99} and the accuracy of the results has been disputed.\cite{dispute} The failure of an independent attempt to detect the injected spins by the spin-e.m.f.\cite{Fil00} supported the conclusion that the level of the spin injection was low. Schmidt {\it et al.}\cite{Sch00} revealed the basic obstacle for the spin injection from a metal into a semiconductor in terms of the conductivity mismatch between the F and N conductors, their conclusions being in agreement with the equations derived in different studies.\cite{vS87,HZ97} Using semimagnetic semiconductors as spin aligners instead of metallic ferromagnets, as proposed in Ref.~\onlinecite{Oes99}, eliminates the mismatch. Indeed, efficient low-temperature spin injection from such sources has been already achieved.\cite{semimagn,Jon00} Another example of the spin injection in semiconductor systems provides a recent theory of the magnetic/nonmagnetic $p-n$-junctions revealing the potentialities of the bipolar spin transport.\cite{ZFDS02} Nevertheless, metallic ferromagnets still remain indispensable sources for the room temperature operating devices.

This seemingly hopeless prognosis for the metallic sources came into conflict with the efficient spin injection into semiconductors from STM tips\cite{Alv95} and resonant double barriers\cite{Oh98}, and also with the injection of hot electrons through Schottky barriers.\cite{Mon98} A solution of this controversy was proposed in the previous paper of the present author.\cite{R00} It was shown that if a non-resistive F-N-interface is replaced by a {\it resistive spin-selective contact}, efficient spin injection from a metallic spin aligner into a semiconductor can be achieved. Under these conditions the spin injection coefficient $\gamma$ is controlled by the competition of three resistances, $r_F$, $r_N$, and $r_c$. Here $r_F\approx L_F/\sigma_F$ and $r_N\approx L_N/\sigma_N$ are the effective spin resistances of the ferromagnet and the normal conductor, respectively, $\sigma_F$ and $\sigma_N$ being their conductivities, $L_F$ and $L_N$ being the spin diffusion lengths in these materials, and $r_c$ is the resistance of a tunnel or a Schottky contact. If $r_c\agt r_F, r_N$, then $\gamma$ is controlled mostly by the spin-selective contact, while the magnitude of the ratio $r_F/r_N$ is of no major importance. However, if $r_c=0$ then $\gamma\sim r_F/r_N$.  When both F and N are metals, the ratio $r_F/r_N\sim 1$ and large $\gamma$ values can be achieved. However, when F is a metal and N is a semiconductor, then $r_F/r_N\ll 1$ and inevitably $\gamma\ll 1$. The last statement is in a complete agreement with the basic conclusion of Ref.~\onlinecite{Sch00}. 

The theory of Ref.~\onlinecite{R00} proved that the spin injection coefficient is controlled by the larger of the resistances in the F-N-junction consisting of a ferromagnetic source, a spin-selective contact, and a normal conductor. Therefore, when F is a metallic source and N is a semiconductor, a resistive spin-selective contact is absolutely necessary to achieve an efficient spin injection. Results of the recent experiments performed by several independent experimental groups are in agreement with this prediction.\cite{Plo01,HJ01,Jon01,Saf01,Weiss02,Jap02,Samarth02} Using resistive contacts turned out advantageous even for all-metallic systems.\cite{Dutch02}  A special type of a contact aimed in even more efficient spin injection have been proposed.\cite{BDF02} Therefore, a theory based on a consistent description of the spin injection in systems including spin-selective contacts is needed. There are several recent papers focused in this problem.\cite{HZ97,R00,FJ01,SS01,R02}

There are two basic types of the electrical measurements that are used for detecting spin injection. The first one is a spin-valve experiment, i.e., a change in the resistance of a F-N-F-junction when the F electrodes are switched between the parallel and antiparallel magnetization directions (the ``conventional" or ``classic" geometry in terms of the Dutch group\cite{JedW}). The second technique measures the spin-e.m.f. and depending on some details of the configuration is termed as a potentiometric or diode geometry by Johnson\cite{HJ01} and as a nonlocal geometry by the Dutch group.\cite{JedW} Optical technique for detecting the non-equilibrium spin-polarization of free carriers\cite{MZ84,KA98} is usually considered as especially reliable and has been applied in a number of recent studies.\cite{Plo01,Jon01,Saf01} 

In the present paper a diffusion theory of F-N- and F-N-F-junctions, that is a useful toy model in the spin injection theory, is developed. The paper includes a multitude of new results and provides a unified approach to the number of mutually related phenomena. It is based on the standard system of equations that are well known, and calculations in essence are elementary. However, they are rather cumbersome and even some of the final results are bulky. Apparently, it is the reason why there is no agreement between the published data, and, hence, there is no consensus about the final results. I show that it is convenient to choose the spin-injection coefficients $\gamma$'s of different contacts as the basic variables (being dependent on the numerous parameters of the system) and to derive a system of equations for them. Such an approach allows (i) to transform the system of equations to a compact form, (ii) to reduce the number of the equations to a minimum, (iii) to receive the final results in an explicit form, and (iv) to establish new regularities. Because in the framework of the ``$\gamma$-technique" the calculations are well organized, the results are reliable and I expect that this technique can be also applied to different systems and geometries and will simplify calculations substantially. 

I have compared my results with a number of the results published previously for different limit cases and presented in various notations. The agreement is emphasized whenewer it was esrablished while the controversies existing in the literature are not disputed.

There are two basic limitations of the theory. First, it is restricted to the diffusion approximation and does not consider neither the Boltzmann approach nor the ballistic transport.\cite{ball} Second, the effect of the spin-orbit interaction in the bulk is neglected. This interaction results in an independent mechanism of the coupling between the electrical current and the electron spin polarization that has been already discussed by different authors.\cite{Rus,West} Because the paper is focused on the theoretical aspects of the problem, I do not discuss the parameters of the specific materials especially as it has been recently done by Fert and Jaffr\`es. \cite{FJ01}

The paper is organized as follows. In Secs.~\ref{sec:basic} and \ref{sec:FN} the basic notations are introduced and the technique based on deriving the equations for $\gamma$'s is illustrated as applied to a F-N-junction. An explicit expression for the resistance of the junction is presented. In Sec.~\ref{sec:FNF} a theory of a F-N-F-junction is developed under rather general assumptions. By deriving and solving the system of two equations for $\gamma$'s the dependence of the spin injection on different parameters is found and analyzed. A direct relation between the spin-valve effect and $\gamma$'s of both contacts is established, and an explicit expression for the resistance of the junction is presented. It is shown that the non-equilibrium part of the resistance related to the spin injection is always positive if both contacts are spin conserving. This equation for the resistance is of current interest especially because of the recent experiment in which it has been measured.\cite{Sch01} In Sec.~\ref{sec:Poi} the role of the Poisson equation in the theory of spin injection is discussed, and a theory of spin-e.m.f. is developed. Division of the parameters into screening dependent and screening independent is clarified. In Sec.~\ref{sec:impedance} the complex impedance of a F-N-junction is found and it is shown that the frequency dependence of the diffusion capacitance and the resistance can become a tool for non-destructive measuring the spin-relaxation times and some different parameters of the junction. In Sec.~\ref{sec:non-cons} the effect of spin non-conservation in a contact on the properties of a F-N-junction is investigated. It is shown that spin non-conservation results in a negative contribution to the junction resistance, and when the spin non-conservation is strong enough even the sign of the non-equilibrium resistance can change. In Sec.~\ref{sec:concl} the basic results are summarized.

\section{Basic equations}              
\label{sec:basic}

Consider first a stationary (dc) one-dimensional flow of electrons across a device consisting of two ferromagnetic conductors, in the regions $x<0$ and $x>d$, and a paramagnetic conductor, in the region $0<x<d$. The N conductor can be both metallic or semiconducting. The F and N regions are connected by tunnel or Schottky contacts.

 In the linear approximation in the total current $J$, the electrical currents $j_{\uparrow,\downarrow}(x)$ carried by up- and down-spins can be written in terms of the space derivatives of the electrochemical potentials 
$\zeta_{\uparrow,\downarrow}(x)$
\begin{equation}
j_{\uparrow,\downarrow}(x)=
\sigma_{\uparrow,\downarrow}\nabla\zeta_{\uparrow,\downarrow}(x)
\label{eq1.1}
\end{equation}
that are related to the non-equilibrium parts $n_{\uparrow,\downarrow}(x)$ of the electron concentrations by the equations 
\begin{equation}
\zeta_{\uparrow,\downarrow}(x)=
(eD_{\uparrow,\downarrow}/\sigma_{\uparrow,\downarrow})n_{\uparrow,\downarrow}(x)-\varphi_F(x).
\label{eq1.2}
\end{equation}
Here $D_{\uparrow,\downarrow}$ and $\sigma_{\uparrow,\downarrow}$ are diffusion coefficients and conductivities, respectively, of up- and down-spin electrons, and electron charge is chosen as $(-e)$, $e>0$. Because of Einstein relations, $D_{\uparrow,\downarrow}$ and 
$\sigma_{\uparrow,\downarrow}$ are related to the temperature dependent densities of states
$\rho_{\uparrow,\downarrow}=\partial n_{\uparrow,\downarrow}/\partial(e\zeta_{\uparrow,\downarrow})$ as
\begin{equation}
e^2D_{\uparrow,\downarrow}=
\sigma_{\uparrow,\downarrow}/\rho_{\uparrow,\downarrow}.
\label{eq1.3}
\end{equation}
The continuity equation in ferromagnets is
\begin{equation}
\nabla j_\uparrow(x)=
{{e^2\rho_\uparrow\rho_\downarrow}\over{\rho_\uparrow+\rho_\downarrow}}
{{\zeta_\uparrow(x)-\zeta_\downarrow(x)}\over{\tau_s^F}},
\label{eq1.4}
\end{equation}
where $\tau^F_s$ is the spin relaxation time. The factor 
$\zeta_\uparrow(x)-\zeta_\downarrow(x)$ ensures the cancelation of the spin relaxation under the conditions of a local spin equilibrium [when 
$\zeta_\uparrow(x)=\zeta_\downarrow(x)$], while the factor
$\rho_\uparrow\rho_\downarrow/(\rho_\uparrow+\rho_\downarrow)$ plays a role of an effective density of states providing the transformation of the difference in the electrochemical potentials into a non-equilibrium concentration.\cite{relax} With a similar equation for $j_\downarrow(x)$, the charge conservation equation reads as
\begin{equation}
j_\uparrow(x)+j_\downarrow(x)=J={\rm const}.
\label{eq1.5}
\end{equation}
Electron concentrations $n_{\uparrow,\downarrow}(x)$ and the electrical potential $\varphi_F(x)$ are related by the Poisson equation that is not written down here because one of the goals of this paper is to establish which physical quantities can be found without using that equation explicitly and which ones depend on it essentially.

It is convenient to use equations symmetrical in the up- and down-spins and for this purpose to define in the both F regions the functions
\begin{equation}
\zeta_F(x)=\zeta_\uparrow(x)-\zeta_\downarrow(x), 
~~Z_F(x)=[\zeta_\uparrow(x)+\zeta_\downarrow(x)]/2
\label{eq1.6}
\end{equation}
and also the ``spin" current, i.e., the difference of the spin polarized currents
\begin{equation}
j_F(x)=j_\uparrow(x)-j_\downarrow(x).
\label{eq1.7}
\end{equation}
Then the spin injection coefficient is
\begin{equation}
\Gamma_F(x)=j_F(x)/J.
\label{eq1.8}
\end{equation}
It follows from Eqs.~(\ref{eq1.1}) and (\ref{eq1.5}) that $\Gamma_F(x)$ is related to $\zeta_F(x)$ by the equation
\begin{equation}
\Gamma_F(x)=
2(\sigma_\uparrow \sigma_\downarrow/\sigma_F)\nabla\zeta_F(x)/J+
\Delta\sigma/\sigma_F,
\label{eq1.9}
\end{equation}
and the usual routine results in the following equations for 
$\zeta_F(x)$ and $Z_F(x)$:
\begin{equation}
\nabla^2\zeta_F(x)=\zeta_F(x)/L_F^2,~~L_F^2=D_F\tau_s^F,
\label{eq1.10}
\end{equation}
\begin{equation}
\nabla Z_F(x)=-(\Delta\sigma/2\sigma_F)\nabla\zeta_F(x)+J/\sigma_F,
\label{eq1.11}
\end{equation}
where the ``bispin" diffusion constant $D_F$ equals
\begin{eqnarray}
D_F&=&(\sigma_\downarrow D_\uparrow +\sigma_\uparrow D_\downarrow)/\sigma_F\nonumber\\
&=&{1\over{e^2}}(\sigma_\uparrow\sigma_\downarrow/\sigma_F)
(\rho_F/\rho_\uparrow\rho_\downarrow)
\label{eq1.11a}
\end{eqnarray}
and
\begin{equation}
\sigma_F=\sigma_\uparrow+\sigma_\downarrow,~~\Delta\sigma=\sigma_\uparrow-\sigma_\downarrow,~~\rho_F=\rho_\uparrow+\rho_\downarrow.
\label{eq1.12}
\end{equation}
Because the parameters of the left and right contacts may not coincide, they will be supplemented in what follows with the subscripts (or superscripts) $L$ and $R$, respectively.

Equations for the N region can be obtained from the above equations by putting 
$\sigma_\uparrow=\sigma_\downarrow=\sigma_N/2$, $\Delta\sigma=0$, $D_\uparrow=D_\downarrow=D_N$:
\begin{eqnarray}
\nabla^2\zeta_N(x)=\zeta_N(x)/L_N^2,&&~~\nabla Z_N(x)=J/\sigma_N\nonumber\\
\Gamma_N(x)=\sigma_N\nabla\zeta_N(x)/2J,&&~~L_N^2=D_N\tau_s^N.
\label{eq1.13}
\end{eqnarray}

The boundary conditions at both contacts may include some additional parameters. It is accepted below that the spin-orbit interaction in the contacts can be neglected, and, therefore, the currents $j_\uparrow$ and $j_\downarrow$ are conserved (see, however, Sec. \ref{sec:non-cons} where the role of the spin non-conservation in a contact is investigated). Because of Eq.~(\ref{eq1.5}), these conditions are equivalent to the conservation of the spin current $j_\uparrow(x)-j_\downarrow(x)$, hence, $j_F(0)=j_N(0)\equiv j(0)$ and
\begin{eqnarray}
\Gamma_F(0)=\Gamma_N(0)\equiv\Gamma_L
\label{eq1.14}
\end{eqnarray}
for the left contact, at $x=0$. Using Eqs.~(\ref{eq1.9}) and (\ref{eq1.13}), this condition can be easily rewritten in an explicit form. A similar equation holds for the right contact, at $x=d$.

All above equations are similar to those of the Shockley theory of the injection through $p-n$-junctions.\cite{Sh49} In that theory the electrochemical potentials of the charge carriers, electrons and holes, are continuous everywhere, in particular, they coincide at both sides of a $p-n$-junction. As applied to the spin injection through resistive contacts a similar condition is fulfilled only if the transparency of the contact, $T$, is large enough, $T\gg(\tau_p/\tau_s)^{1/2}$, where $\tau_p$ and $\tau_s$ are the momentum and spin relaxation times, respectively.\cite{T} However, an efficient spin injection is facilitated by the low contact transparency,\cite{R00} and under these conditions the electrochemical potentials of up- and down-spin electrons are discontinuous at the contacts. For spin-selective contacts with the different conductivities for the up- and down-spin electrons, the boundary condition for the left contact, at $x=0$, reads as 
\begin{equation}
j_{\uparrow,\downarrow}(0)=\Sigma^L_{\uparrow,\downarrow}(\zeta^N_{\uparrow,\downarrow}(0)
-\zeta^F_{\uparrow,\downarrow}(0)).
\label{eq1.15}
\end{equation}
Here $\Sigma^L_\uparrow$ and $\Sigma^L_\downarrow$ are the contact conductivities for the up- and down-spin electrons, respectively. A similar equation should be written for the right contact, at $x=d$. In the symmetric variables of Eq.~(\ref{eq1.6}), equation (\ref{eq1.15}) takes the form
\begin{equation}
\zeta_N(0)-\zeta_F(0)= 2r_c^L(\Gamma_L-\Delta\Sigma_L/\Sigma_L)J,
\label{eq1.16}
\end{equation}
\begin{equation}
Z_N(0)-Z_F(0)=r_c^L[1-(\Delta\Sigma_L/\Sigma_L)\Gamma_L]J.
\label{eq1.17}
\end{equation}
Here 
\begin{equation}
\Delta\Sigma_L=\Sigma^L_\uparrow-\Sigma^L_\downarrow,~
 \Sigma_L=\Sigma^L_\uparrow+\Sigma^L_\downarrow,~
r_c^L={\Sigma_L}/{4\Sigma^L_\uparrow\Sigma^L_\downarrow}, 
\label{eq1.18}
\end{equation}
and $r_c^L$ plays a role of the effective contact resistance. Similar equations hold for the right contact.

It follows from Eqs.~(\ref{eq1.9}), (\ref{eq1.10}), (\ref{eq1.13}), and 
(\ref{eq1.16}) that the equations for $\Gamma_F(x)$ and $\zeta_F(x)$, and also for $\Gamma_N(x)$ and $\zeta_N(x)$, completely separate from the equations for 
$Z_F(x)$ and $Z_N(x)$. This property that is valid only for the spin-conserving contacts in the dc regime, cf. Secs.~\ref{sec:impedance} and \ref{sec:non-cons}, simplifies essentially the following calculations.

\section{F-N-junction}              
\label{sec:FN}

A F-N-junction with a spin-selective contact at $x=0$ provides the simplest model for the specific features of the spin injection to manifest themselves, and it has been discussed already in a number of papers.\cite{vS87,HZ97,R00} For semi-infinite F and N regions, the functions $\zeta_F(x)$ and $\zeta_N(x)$ decay exponentially as
\begin{equation}
\zeta_F(x)=\zeta_F\exp(x/L_F),~~\zeta_N(x)=\zeta_N\exp(-x/L_N), 
\label{eq2.1}
\end{equation}
and Eqs.~(\ref{eq1.9}) and (\ref{eq1.13}) permit one to express the integration constants $\zeta_F$ and $\zeta_N$ through the spin injection coefficient, $\Gamma$. In this section and everywhere below it will be denoted as $\gamma$ as applied to a F-N-junction with semi-infinite conductors on its left and right hand sides. Then
\begin{equation}
\zeta_F=2r_F(\gamma-\Delta\sigma/\sigma_F)J,~~\zeta_N=-2r_N\gamma J, 
\label{eq2.2}
\end{equation}
where
\begin{equation}
r_N=L_N/\sigma_N,~~r_F=\sigma_F L_F/4\sigma_\uparrow\sigma_\downarrow 
\label{eq2.3}
\end{equation}
are the effective resistances of the N and F regions, respectively. The index $L$ is omitted everywhere in this section.

Substituting $\zeta_F$ and $\zeta_N$ from Eq.~(\ref{eq2.2}) into Eq.~(\ref{eq1.16}), one finds readily the self-consistency equation for $\gamma$:
\[-(r_N+r_F)\gamma+r_F\Delta\sigma/\sigma_F=r_c(\gamma-\Delta\Sigma/\Sigma).\]
 Its solution reads
\begin{equation}
\gamma=[r_c(\Delta\Sigma/\Sigma)+r_F(\Delta\sigma/\sigma_F)]/r_{FN}~, 
\label{eq2.4}
\end{equation}
where
\begin{equation}
r_{FN}=r_F+r_c+r_N 
\label{eq2.5}
\end{equation}
can be considered as an effective resistance of the F-N-junction. This equation is equivalent to Eq.~(3.19) by Hershfield and Zhao,\cite{HZ97} and in the limit $r_c=0$ also to the result by van Son {\it et al.}\cite{vS87}

The above derivation of Eq.~(\ref{eq2.4}) exemplifies application of the $\gamma$-{\it technique} based on eliminating the external parts of the device by {\it choosing the spin injection coefficients} $\Gamma$ (or $\gamma$) {\it at the boundaries as the basic variables} with subsequent deriving and solving the self-consistency equations for these $\Gamma$'s. It simplifies the calculations dramatically and will be employed everywhere in what follows.

Eq.~(\ref{eq2.4}) displays in a simple way the basic regularities related to the spin injection. For a metallic F conductor and a semiconducting N conductor usually $r_F\ll r_N$, while for a good Ohmic contact between them $r_c= 0$. Hence, $\gamma\sim r_F/r_N\ll 1$. Therefore, when $r_c=0$ the conductivity mismatch becomes an obstacle for an efficient spin injection.\cite{Sch00} Eq.~(\ref{eq2.4}) suggests that there are several ways to remedy the problem: (i) using metallic ferromagnets in combination with spin-selective resistive contacts, $r_c \agt r_N, r_F$,\cite{R00} (ii) using half metallic spin emitters\cite{PM01,Tal} because for them either $\sigma_\uparrow$ or $\sigma_\downarrow$ is very small and, therefore, $r_F$ is anomalously large [cf. Eq.~(\ref{eq2.3})],\cite{LaB01} or (iii) using semimagnetic semiconductors as spin aligners because for them $r_F\sim r_N$.\cite{Oes99,semimagn,Jon00} All these options are currently investigated experimentally.

The standard approach for calculating the resistance of a junction, $R$, is based on the fact that far from the junction, $\vert x\vert\gg L_F, L_N$, where $n_{\uparrow,\downarrow}(x)\approx 0$, the electrochemical potentials $\zeta_{\uparrow,\downarrow}(x)$ and the electrical potential $\varphi_F(x)$ show the same behavior, $\zeta_{\uparrow,\downarrow}(x)\approx -\varphi_F(x)$ as it follows from Eq.~(\ref{eq1.2}). Therefore, the potential drop across the specimen equals to the variance of $Z(x)$ across it. Integrating the equations (\ref{eq1.11}) and (\ref{eq1.13}) for $Z_F(x)$ and $Z_N(x)$ results in
\begin{eqnarray}
Z_F(x)&=&-(\Delta\sigma/2\sigma_F)\zeta_F(x)+Jx/\sigma_F+C_F, \nonumber \\
Z_N(x)&=&Jx/\sigma_N+C_N, 
\label{eq2.6}
\end{eqnarray}
where $C_F$ and $C_N$ are the integration constants. After subtracting the nominal Ohmic potential drop at the segments $(x_L, 0)$ and $(0, x_R)$, $x_L$ and $x_R$ being the left and the right ends of the device, one finds from $Z_N(x_R)-Z_F(x_L)$ that $RJ=C_N-C_F$. This latter difference can be derived from Eq.~(\ref{eq1.17}). With a proper account of Eqs.~(\ref{eq2.2})-(\ref{eq2.5}), one gets after a simple algebra:
\begin{eqnarray}
 R(\gamma)&=&R_{\rm eq}+R_{\rm n-eq}(\gamma),~~R_{\rm eq}=\Sigma^{-1},\nonumber \\
R_{\rm n-eq}(\gamma)&=&[r_F(\Delta\sigma/\sigma_F)^2
+r_c(\Delta\Sigma/\Sigma)^2]-\gamma^2r_{FN}.
\label{eq2.7}
\end{eqnarray}
This result recovers Eq.~(21) of Ref.\onlinecite{R00}, and for $r_c=0$ it is in agreement with the result of Ref.\onlinecite{vS87}.

It is a remarkable property of Eq.~(\ref{eq2.7}) that it {\it relates the resistance} $R$ {\it to} $\gamma$.

In absence of non-equilibrium spins, when $L_F,L_N\rightarrow 0$, only the first term, $R_{\rm eq}$, survives in Eq.~(\ref{eq2.7}) while the two terms making $R_{\rm n-eq}(\gamma)$ cancel.\cite{resist} The first term in $R_{\rm n-eq}$ is always positive and can be identified as the Kapitza resistance originating from the conversion of the spin flows near the contact. The second term in $R_{\rm n-eq}$, the injection conductivity, is negative and directly related to the spin injection.\cite{R00} There are good reasons for dividing $R_{\rm n-eq}$ into these two parts. Indeed, it will be shown in Sec.~\ref{sec:FNF} that the spin valve effect in F-N-F structures originates completely from the injection conductivity.

The nonequilibrium resistance of Eq.~(\ref{eq2.7}), $R_{\rm n-eq}$, includes two competing terms. Substituting $\gamma$ from Eq.~(\ref{eq2.4}) into Eq.~(\ref{eq2.7}) results in an explicit equation for $R_{\rm n-eq}$:
\begin{eqnarray}
 R_{\rm n-eq}&=&{1\over{r_{FN}}}\{ r_N\left[r_c(\Delta\Sigma/\Sigma)^2
+r_F(\Delta\sigma/\sigma_F)^2 \right]\nonumber \\
&+&r_c r_F\left[(\Delta\Sigma/\Sigma)-(\Delta\sigma/\sigma_F)\right]^2 \}. 
\label{eq2.8}
\end{eqnarray}
It is seen from Eq.~(\ref{eq2.8}) that $R_{\rm n-eq}$ {\it is always positive}, i.e., the spin injection enlarges the junction resistance as it has been already stated in Ref.\onlinecite{R00}. 

It follows foom Eq.~(\ref{eq2.7}) that there is a critical difference between the spin injection through resistive contacts and the bipolar injection across a $p-n$-junction. The conductivity of a $p-n$-junction comes from the generation or recombination of non-equilibrium carriers while a large resistance of the depletion layer plays no essential role. On the contrary, for a F-N-junction the equilibrium resistance $R_{\rm eq}$ should be added to $R_{\rm n-eq}$. It is only in the limit $r_c=0$ that $R^{-1}_{\rm n-eq}=(r_F^{-1}+r_N^{-1})(\sigma_F/\Delta\sigma)^2$ takes a form similar to the Shockley conductivity. However, this limit is only of a limited interest as applied to the semiconductor spin-injection devices because of small $\gamma$. It is also worth mentioning that the conductivities of the different elements of a F-N-junction enter into Eq.~(\ref{eq2.8}) in a rather peculiar way. Therefore, the prospects for finding simple ``equivalent schemes" for spin injection devices do not seem promising.

The above results were derived without using the Poisson equation (or the quasineutrality condition instead of it) explicitly. In this connection some comments are needed. First, using $\nabla Z(x)$ instead of $\nabla\varphi(x)$ is justified only as applied to a two terminal resistance when the potential drop at the scale large compared to $L_F$ and $L_N$ is relevant. The ``equivalent" electrical field\cite{VF93} $\nabla Z(x)$ differs essentially from the ``genuine" electrical field $-\nabla\varphi(x)$, see Eq.~(\ref{eq4.2}) below. Second, for calculating the impedance of a F-N-junction the Poisson equation (or the quasineutrality condition) should be used explicitly, cf. Sec.~\ref{sec:impedance}. Third, the linear dependence of $Z(x)$ and $\varphi(x)$ on $x$ for $\vert x\vert\gg L_F,L_N$ [cf. Eq.~(\ref{eq2.6})] is based implicitly on the assumption of a 3D screening of the electrical field in the leads. It is not valid for low-dimensional conductors with a strong size quantization, hence, they require a special consideration, see Sec.~\ref{sec:Poi} below.

\section{F-N-F-junction}              
\label{sec:FNF}

For a F-N-F-junction already discussed in Sec.~\ref{sec:basic}, the solutions for $\zeta^L_F(x)$ and $\zeta^R_F(x)$ in the two infinite regions, $x<0$ and $x>d$, respectively, are described by Eq.~(\ref{eq2.1}) with only a little change:
\begin{equation}
\zeta^L_F(x)=\zeta_F^L\ e^{x/L_F^L},~\zeta_F^R(x)=\zeta_F^R e^{(d-x)/L_F^R}.
\label{eq3.1}
\end{equation}
The indeces $L$ and $R$ specify the left and the right ferromagnets, respectively. Eq.~(\ref{eq1.9}) allows to express the integration constants, $\zeta_F^L$ and $\zeta^R_F$, in terms of the coefficients of the spin injection, $\Gamma_L$ and $\Gamma_R$, across the left and the right contact, respectively:
\begin{equation}
\zeta^L_F=2r_F^L(\Gamma_L-{{\Delta\sigma_L}\over{\sigma_F^L}})J,~
\zeta^R_F=-2r_F^R(\Gamma_R-{{\Delta\sigma_R}\over{\sigma_F^R}})J.
\label{eq3.2}
\end{equation}
In a similar way, one finds the electrochemical potential $\zeta_N(x)$ in the central region in terms of the same spin-injection coefficients $\Gamma_L$ and $\Gamma_R$:
\begin{eqnarray}
\zeta_N(x)&=&2r_N\{\Gamma_R\cosh(x/L_N)\nonumber \\
&-&\Gamma_L\cosh[(d-x)/L_N)]\}J/\sinh(d/L_N).
\label{eq3.3}
\end{eqnarray}

The spin current conservation law of Eq.~(\ref{eq1.14}) is already satisfied and manifests itself in the fact that $\Gamma_L$ and $\Gamma_R$ in Eqs.~(\ref{eq3.2}) and (\ref{eq3.3}) do not bear the indeces F and N. Eq.~(\ref{eq1.16}) and a similar equation for the right contact result, after some algebra, in the system of two equations of the $\gamma$-technique for $\Gamma_L$ and $\Gamma_R$:
\begin{eqnarray}
r_{FN}^L(d)\Gamma_L-\left\{r_N/\sinh(d/L_N)\right\}\Gamma_R&=&r_{FN}^L\gamma_L,\nonumber \\
-\left\{r_N/\sinh(d/L_N)\right\}\Gamma_L+r_{FN}^R(d)\Gamma_R&=&r_{FN}^R\gamma_R.
\label{eq3.4}
\end{eqnarray}
Here $\gamma_L$ and $\gamma_R$ are the spin injection coefficients, Eq.~(\ref{eq2.4}), for the identical contacts but surrounded by infinite F and N leads, $r_{FN}^L$ and $r_{FN}^R$ are the effective resistances of these contacts, Eq.~(\ref{eq2.5}), and
\begin{equation}
r_{FN}^{L(R)}(d)=r_F^{L(R)}+r_c^{L(R)}+r_N\coth(d/L_N).
\label{eq3.5}
\end{equation}
{\it The system of two equations (\ref{eq3.4}) solves completely the problem of the spin injection across a F-N-F-junction}. 

\subsection{Spin injection}              
\label{sec:inj}

The solution of the equations (\ref{eq3.4}) is:
\begin{eqnarray}
\Gamma_L=\left[\gamma_Lr^L_{FN}r^R_{FN}(d)+\gamma_Rr_Nr^R_{FN}/\sinh(d/L_N)\right]/{\cal D},\nonumber \\
\Gamma_R=\left[\gamma_Rr^R_{FN}r^L_{FN}(d)+\gamma_Lr_Nr^L_{FN}/\sinh(d/L_N)\right]/{\cal D},
\label{eq3.6}
\end{eqnarray}
where $\cal D$ is the determinant of the system 
\begin{eqnarray}
{\cal D}&=&(r_F^L+r_c^L)(r_c^R+r_F^R)\nonumber \\
&+&r_N^2+r_N(r_F^L+r_c^L+r_c^R+r_F^R)\coth(d/L_N).
\label{eq3.7}
\end{eqnarray}
$\Gamma_L$ and $\Gamma_R$ determine the spin injection coefficients across both contacts, while Eqs.~(\ref{eq3.3}) and (\ref{eq1.13}) provide the distribution of the electrochemical potential $\zeta_N(x)$ and of the spin current $\Gamma_N(x)$ inside the N-region.

It is instructive to consider several limit cases, especially because the criteria of their validity are not obvious.

\subsubsection{Wide N region}
\label{sec:wideN}

If $d/L_N\gg 1$, then $r_{FN}^{L(R)}(d)\approx r_{FN}^{L(R)}$ and 
${\cal D}\approx r_{FN}^Lr_{FN}^R$ with the exponential accuracy, and the second terms in the nominators of $\Gamma_L$ and $\Gamma_R$ can be omitted. Finally
\begin{equation}
\Gamma_L\approx\gamma_L,~~ \Gamma_R\approx\gamma_R .
\label{eq3.8}
\end{equation}
Therefore, the contacts are completely decoupled.

\subsubsection{Narrow N region}
\label{sec:narN}

If the criteria $d/L_N\ll 1,~ r_N/r_c^{L(R)},~ r_N/r_F^{L(R)}$ are fulfilled, then the spin relaxation in the N-region plays only a minor role, and $\Gamma_L$ and $\Gamma_R$ nearly coincide even when $\gamma_L\neq\gamma_R$. Indeed, if follows from Eqs.~(\ref{eq3.5}) -- (\ref{eq3.7}) that 
\begin{equation}
\Gamma\equiv\Gamma_L=\Gamma_R=\left( r_{FN}^L\gamma_L+r_{FN}^R\gamma_R\right)/r_{LNR}
\label{eq3.9}
\end{equation}
where
\begin{equation}
r_{LNR}=r_F^L+r_c^L+d/\sigma_L+r_c^R+r_F^R.
\label{eq3.10}
\end{equation}
Therefore, under these conditions the N region takes the control over the injection from both contacts keeping it at an equal level. Eqs.~(\ref{eq3.9}) and (\ref{eq3.10}) coincide with Eq.~(22) of Ref.~\onlinecite{R00}.

It is seen from the comparison of Eqs.~(\ref{eq3.3}) and (\ref{eq3.9}) that for a parallel magnetization alignment of two identical F conductors ($\gamma_L=\gamma_R$) the spin injection coefficient $\Gamma$ is large but the concentration of nonequilibrium spins in the N region, that is proportional to $\zeta_N$, is small in the parameter $d/L_N$. Just opposite, for an antiparallel alignment ($\gamma_L=-\gamma_R$) the spin injection coefficient is small while the concentration of nonequilibrium spins remains finite when $d/L_N\rightarrow 0$ as one can check using Eqs.~(\ref{eq3.3}) and (\ref{eq3.13}), see below.

\subsubsection{High-resistance contacts}
\label{sec:highRes}

When both the criterion $r_c^{L(R)}\gg r_N\coth(d/L_N)$ and the criterion $r_c^{L(R)}\gg r_N,~r_F^{L(R)}$ are satisfied, the contact resistances $r_c^{L(R)}$ dominate in all equations. Therefore
\begin{equation}
\Gamma_{L(R)}=[r_c^{L(R)}/(r_c^L+r_c^R)](\Delta\Sigma_{L(R)}/\Sigma_{L(R)}).
\label{eq3.11}
\end{equation}
The contact resistances completely control the injection regime. It is this regime that has been proposed in Ref.~\onlinecite{R00} to remedy the problem of the conductivity mismatch. The recent experimental data\cite{Plo01,HJ01,Jon01,Saf01,Weiss02,Jap02,Samarth02} are in a reasonable agreement with the theoretical expectations.  

One can easily see that Eqs.~(\ref{eq3.9}) and (\ref{eq3.11}) are incompatible. This fact indicates that the parameter region
\begin{equation}
r_c^{L(R)}\approx r_NL_N/d
\label{eq3.12}
\end{equation}
is critical for the spin injection regime that switches between the N-region controlled and the contact controlled limits. Physical implications of this criterion have been discussed in many details in the recent paper by Fert and 
Jaffr\`es.\cite{FJ01} In particular, Eq.~(22) of Ref.~\onlinecite{R00} found under the conditions $\Gamma(x)=$const inside the N-region is applicable only when $w/L_N\ll r_N/r_c^{L(R)}$. 

\subsubsection{Symmetric geometry}
\label{sec:symGeo}

If the parameters of both ferromagnetic electrodes and both contacts are identical, there remain two options of the parallel, $\Delta\sigma_L=\Delta\sigma_R$ and $\Delta\Sigma_L=\Delta\Sigma_R$, and the antiparallel, $\Delta\sigma_L=-\Delta\sigma_R$ and $\Delta\Sigma_L=-\Delta\Sigma_R$, alignment of the magnetization of the electrodes. In the symmetric geometry the determinant $\cal D$ factorizes as
\[{\cal D}=[r_F+r_c+r_N\tanh(d/2L_N)][r_F+r_c+r_N\coth(d/2L_N)],\]
and Eq.~(\ref{eq3.6}) can be simplified for the both alignments
\begin{eqnarray}
\Gamma_L^{(p)}&=&\Gamma_R^{(p)}=\gamma_Lr_{FN}/[r_F+r_c+r_N\tanh(d/2L_N)]
\nonumber \\
\Gamma_L^{(ap)}&=&-\Gamma_R^{(ap)}=\gamma_Lr_{FN}/[r_F+r_c+r_N\coth(d/2L_N)].
\label{eq3.13}
\end{eqnarray}
It is seen that $\Gamma_L^{(p)}>\gamma_L >\Gamma_L^{(ap)}$. If the N region is narrow enough, so that $d/2L_N\ll 1$ and $2r_NL_N/d\gg r_F,r_c$, then $\Gamma_L^{(ap)}\ll\gamma_L$, in accordance with the result following from Eq.~(\ref{eq3.9}) for the antiparallel alignment. When $r_c=0$ and $d/2L_N\ll 1$, Eq.~(7) of Ref.\onlinecite{Sch00} can be recovered from Eqs.~(\ref{eq3.13}).

\subsection{Resistance: general equation}
\label{sec:resist}              

To find the resistance $\cal R$ of a F-N-F-junction one has to write down the equations for $Z_F^L(x)$, $Z_N(x)$, and $Z_F^R(x)$. It follows from Eqs.~(\ref{eq1.11}) and (\ref{eq1.13}) that
\begin{eqnarray}
Z_F^L(x)&=&-(\Delta\sigma_L/2\sigma_F^L)\zeta_F^L(x)+Jx/\sigma_F+C_L,\nonumber \\
Z_N(x)&=&Jx/\sigma_N+C_N,\nonumber \\
Z_F^R(x)&=&-(\Delta\sigma_R/2\sigma_F^R)\zeta_F^R(x)+Jx/\sigma_F+C_R,
\label{eq3.14}
\end{eqnarray}
where $C_L$, $C_N$, and $C_R$ are integration constants. The arguments similar to those of Sec.~\ref{sec:FN} allow to relate them to $\cal R$:
\begin{equation}
J{\cal R}=(C_R-C_L) -d(1/\sigma_N-1/\sigma_F^R).
\label{eq3.15}
\end{equation}
The nominal Ohmic resistance of the N-conductor and the both F-conductors is subtracted from $\cal R$. Applying Eq.~(\ref{eq1.17}) to the $L$ and $R$ contacts one finds $C_N-C_L$ and $C_R-C_N$, and then 
\begin{eqnarray}
{\cal R}&=&r_F^L(1-\Gamma_L\Delta\Sigma_L/\Sigma_L)+r_F^R(1-\Gamma_R\Delta\Sigma_R/\Sigma_R)
\nonumber \\
&-&(\Delta\sigma_L/2\sigma_F^L)\zeta_F^L/J+(\Delta\sigma_R/2\sigma_F^R)\zeta_F^R/J.
\label{eq3.16}
\end{eqnarray}
Plugging the expressions for $\zeta_F^L$ and $\zeta_F^R$ from Eq.~(\ref{eq3.2}) into this equation results in the resistance $\cal R$ expressed in terms of $\Gamma_L$ and $\Gamma_R$. Similar to Eq.~(\ref{eq2.7}), $\cal R$ can be split into the equilibrium and non-equilibrium parts, ${\cal R}_{\rm eq}$ and 
${\cal R}_{\rm n-eq}$, respectively:
\begin{equation}
{\cal R}={\cal R}_{\rm eq}+{\cal R}_{\rm n-eq},
\label{eq3.16a}
\end{equation}
where
\begin{equation}
{\cal R}_{\rm eq}=\Sigma_L^{-1}+\Sigma_R^{-1}
\label{eq3.16b}
\end{equation}
makes the total resistance in the limit $L_F^{L(R)}, L_N\rightarrow 0$, i.e., when the concentrations of the non-equilibrium spins vanish in both F-regions and in the N-region. Finally

\end{multicols}
\widetext 

\begin{equation}
{\cal R}_{\rm n-eq}(\Gamma_L,\Gamma_R)=
\left[r_F^L(\Delta\sigma_L/\sigma_F^L)^2
+r_F^R(\Delta\sigma_R/\sigma_F^R)^2\right]
+\left[r_c^L(\Delta\Sigma_L/\Sigma_L)^2
+r_c^R(\Delta\Sigma_R/\Sigma_R)^2\right]
-\left(r_{FN}^L\gamma_L\Gamma_L+r_{FN}^R\gamma_R\Gamma_R\right).
\label{eq3.17}
\end{equation}
With $\Gamma_L$ and $\Gamma_R$ defined by Eqs.~(\ref{eq3.6}), this equation provides a general expression for the resistance of a F-N-F-junction. Its explicit form in terms of $\gamma_L$ and $\gamma_R$ is
\begin{eqnarray}
{\cal R}_{\rm n-eq}(\gamma_L,\gamma_R)&=&
\left[r_F^L(\Delta\sigma_L/\sigma_F^L)^2
+r_F^R(\Delta\sigma_R/\sigma_F^R)^2\right]
+\left[r_c^L(\Delta\Sigma_L/\Sigma_L)^2
+r_c^R(\Delta\Sigma_R/\Sigma_R)^2\right]\nonumber\\
&-&[\gamma_L^2(r_{FN}^L)^2r_{FN}^R(d)
+\gamma_R^2(r_{FN}^R)^2r_{FN}^L(d)]/{\cal D}
-2\gamma_L\gamma_Rr_{FN}^Lr_{FN}^Rr_N/{\cal D}\sinh(d/L_N).
\label{eq3.18}
\end{eqnarray}

\begin{multicols}{2}
\noindent
Eq.~(\ref{eq3.18}) is applicable for arbitrary values of the parameters of both contacts.

There is a similarity between Eqs.~(\ref{eq2.7}) and (\ref{eq3.18}). Indeed, they both include the injection conductivity, i.e., the terms quadratic in $\gamma$'s (the spin injection coefficients of two unlimited F-N-junctions). Application of Eq.~(\ref{eq3.18}) for calculating the spin valve effect, see Sec.~\ref{sec:valve} below, demonstrates convincingly how useful this representation is. Eq.~(\ref{eq3.18}), being cumbersome in its general form, essentially simplifies in different limit cases.

If the spin relaxation in the N region can be neglected, $d/L_N\rightarrow 0$, then 
$\Gamma_L=\Gamma_R\equiv\Gamma$ and the last term of Eq.~(\ref{eq3.17}) takes the form $-r_{FNF}\Gamma^2$ as it follows from the comparison with Eq.~(\ref{eq3.9}). There is a striking similarity between this term and the last term of Eq.~(\ref{eq2.7}) for the resistance of a F-N-junction. Therefore, the representation for $R$ and $\cal R$ in terms of $\gamma$'s underscores a close interconnection between the properties of F-N- and F-N-F-junctions.

Quite similar to a F-N-junction (cf. Sec.~\ref{sec:FN}), the non-equilibrium  resistance ${\cal R}_{\rm n-eq}$ is always positive, ${\cal R}_{\rm n-eq}>0$. This point will be discussed in Sec.~\ref{sec:neq} below. 

\subsubsection{Spin valve effect}
\label{sec:valve}

One of the basic parameters of a F-N-F-junction is the switching resistance $\Delta{\cal R}$ also termed as a spin valve effect. It equals to the difference in the resistances of a F-N-F-junction for the antiparallel and parallel alignment of the magnetization of the two F conductors. $\Delta{\cal R}$ is equal to the change in the resistance when $\gamma_R$ (or $\gamma_L$) changes sign. It is obvious from Eq.~(\ref{eq3.18}) that the spin valve effect comes exclusively from its last term that is proportional to the product of $\gamma_L$ and $\gamma_R$ while the different terms including $\gamma_L$ and $\gamma_R$ are proportional to their squares and hence do not change when one of $\gamma$'s changes sign. Therefore:
\begin{equation}
\Delta{\cal R}={\cal R}_{\uparrow\downarrow}-{\cal R}_{\uparrow\uparrow}=
4\gamma_L\gamma_R\cdot{{r_{FN}^Lr_{FN}^Rr_N}\over{{\cal D}\sinh(d/L_N)}}.
\label{eq3.19}
\end{equation}
This result seems to have the same functional form but be twice as large as the result derived by some indirect procedure based on Onsager relations in Ref.\onlinecite{HZ97} [cf. their Eq.~(5.9) written in different notations]. In the limit $d/L_N\rightarrow 0$, Eq.~(\ref{eq3.19}) coincides with Eq.~(25) of Ref.\onlinecite{R00}.

In the symmetric geometry of Sec.~\ref{sec:symGeo}, the explicit form of 
Eq.~(\ref{eq3.19}) is:
\end{multicols}
\widetext 
\begin{equation}
\Delta{\cal R}= 
{{4r_N\left(r_c{{\Delta\Sigma}\over\Sigma}
+r_F{{\Delta\sigma}\over{\sigma_F}}\right)^2}\over 
{[(r_F+r_c)^2+r_N^2]\sinh(d/L_N)+2r_N(r_F+r_c)\cosh(d/L_N)}}
\label{eq3.19a}
\end{equation}
\begin{multicols}{2}
\noindent
This result differs from the equations available in the literature.

Using Eqs.~(\ref{eq2.4}) and (\ref{eq3.7}), one can find from Eq.~(\ref{eq3.19a}) the switching resistance $\Delta{\cal R}$ for different limit cases. E.g., in half-metallic wires the resistances $r_F^{L(R)}$ are controlled by minority carriers and, therefore, are very large. Assuming $r_c^{L(R)}\ll r_F^{L(R)}$, one can check that $\Delta{\cal R}$ increases monotonically with $r_F^L$ and $r_F^R$. In the limit of large $r_F^Lr_F^R$ one finds $\Delta{\cal R}\approx 4r_N/\sinh(d/L_N)$. Strong increase in $\Delta{\cal R}$ for thin samples, $d/L_N\ll 1$,  is in a qualitative agreement with existing theories.\cite{VF93}

\subsubsection{Non-equilibrium resistance}
\label{sec:neq}

The non-equilibrium part of the resistance, Eq.~(\ref{eq3.18}), includes competing positive and negative terms. However, it has been emphasized in Ref.~\onlinecite{R00} that ${\cal R}_{\rm n-eq}>0$, quite similar to Eq.~(\ref{eq2.8}), and this statement holds both for the parallel and antiparallel spin alignments. The algebraic transformation proving this statement is lengthly, and even the final expression is rather bulky. A general expression for ${\cal R}_{\rm n-eq}$ in the form of a sum of obviously positive terms is presented in Appendix. An experimental observation of the positive magnetoresistance originating from ${\cal R}_{\rm n-eq}>0$ has been reported recently.\cite{Sch01} It was detected as a considerable increase in the resistance of a F-N-F-junction in a weak magnetic field aligning the spin-polarized contacts prepared from a dilute semimagnetic semiconductor.

The general Eqs.~(\ref{A1}) and (\ref{A2}) for ${\cal R}_{\rm n-eq}$ can be essentially simplified in different limit cases. E.g., for a system with a symmetric geometry (cf. Sec.~\ref{sec:symGeo}), the parallel spin alignment, and a zero contact resistance
\begin{eqnarray}
{\cal R}_{\rm n-eq}&=&2r_Fr_N\left(\Delta\sigma/\sigma_F\right)^2\nonumber\\&\times&
{{r_N\sinh(d/L_N)+r_F[\cosh(d/L_N)-1]}\over{(r_F^2+r_N^2)\sinh(d/L_N)+2r_Fr_N\cosh(d/L_N)}}.
\label{eq3.20}
\end{eqnarray}
This expression is obviously positive and is even in the factor $\Delta\sigma/\sigma_F$ as one expects because simultaneous reversing the magnetization of both ferromagnetic conductors should not change the resistance of the junction. For $d\gg L_N$, it follows from Eq.~(\ref{eq3.20}) that ${\cal R}_{\rm n-eq}\approx[2r_Fr_N/(r_F+r_N)](\Delta\sigma/\sigma_F)^2$. This resistance is twice as large as the non-equilibrium resistance of a F-N-junction found from Eq.~(\ref{eq2.8}), just as one expects. In the opposite limit of a narrow N-region, $d/L_N\rightarrow 0$, one finds ${\cal R}_{\rm n-eq}/{\cal R}_d\approx(\Delta\sigma/\sigma_F)^2$, where ${\cal R}_d=d/\sigma_N$ is the nominal resistance of the N-region. Therefore, for $\Delta\sigma/\sigma_F\sim 1$ the effect is really large.

\section{Poisson equation}
\label{sec:Poi}

All above equations for the spin injection coefficients and the two-terminal resistances $R$ and $\cal R$ of F-N- and F-N-F-junctions, respectively, were derived without using Eq.~(\ref{eq1.2}) relating the electrical and electrochemical potentials. The Poisson equation for the electrical potential $\varphi(x)$ has also not been used explicitly, as well as the quasineutrality condition widely applied instead of it in the theory of semiconductor devices outside the space charge region. However, the identification of the difference of the integration constants $C_N-C_F$ and $C_R-C_L$ with the potential drop (cf. Secs.~\ref{sec:FN} and \ref{sec:resist}, respectively) involves the concept of the linear in $x$ dependence of the potential $\varphi(x)$ far away from the junction. It is typical of the 3D conductors with the characteristic dimensions large compared to the screening length ${\it l}_{\rm sc}\sim\sqrt{\varepsilon/4\pi e^2\rho}$. More subtle arguments are needed as applied to the two-terminal resistance of 2D and 1D ``quantum" conductors that exhibit a strong size quantization and whose transverse dimensions are small compared to ${\it l}_{\rm sc}$. Those arguments involve restoring the global neutrality due to external ``classical" electrodes and are reminiscent of the Thouless arguments in the theory of the Landauer conductance.\cite{Th81} Meanwhile, the linear in $x$ behavior of $\varphi(x)$ can be violated in low-dimensional conductors at a macroscopic scale large compared to ${\it l}_{\rm sc}$.\cite{KRnon} In the present paper, I do not discuss these specific properties of low-dimensional conductors and accept the screening be three-dimensional with ${\it l}_{\rm sc}\ll L_F,L_N$. Therefore, the quasineutrality approximation is applicable.

However, measuring the two terminal resistance (the ``conventional" or ``classic" geometry) is not the only method for detecting and measuring the spin injection and it suffers from some drawbacks as it has been discussed in Ref.~\onlinecite{JedW}. The different methods include measuring the spin-emf (or, what is the same, the open circuit voltage) at a spin-selective electrode in its different modifications,\cite{JS85,HJ01,JedW} the optical detection of spin-polarized carriers,\cite{Plo01,Jon01,Saf01,MZ84,KA98,Hag98} etc. All these methods are based on separate measuring the two terms entering into the electrochemical potentials of 
Eq.~(\ref{eq1.2}): the concentrations $n_{\uparrow,\downarrow}(x)$ and the electrical potential $\varphi(x)$. 

The quasineutrality condition for the concentrations of non-equilibrium spins, 
$n_\uparrow(x)$ and $n_\downarrow(x)$,
\begin{equation}
n_\uparrow(x)+n_\downarrow(x)=0,
\label{eq4.1}
\end{equation}
together with Eqs.~(\ref{eq1.2}) and (\ref{eq1.3}), results in an equation for the electrical potential $\varphi_F(x)$ in the ferromagnetic regions
\begin{equation}
\varphi_F(x)=-Z_F(x)-{\case 1/2}(\Delta\rho/\rho_F)\zeta_F(z),
\label{eq4.2}
\end{equation}
where $\Delta\rho=\rho_\uparrow-\rho_\downarrow$.
It follows from a similar equation for the N region, $\varphi_N(x)=-Z_N(x)$, that there the ``equivalent" field $\nabla Z_N(x)$ and the electrical field 
$-\nabla\varphi_N(x)$ do coincide. However, in the F regions the fields $\nabla Z_F(x)$ and $-\nabla\varphi_F(x)$ show rather different behavior everywhere inside the spin injection zone, $\zeta_F(x)\neq 0$. 

From these equations and Eq.~(\ref{eq1.17}) one finds the potential drop at the $x=0$ contact:
\begin{eqnarray}
\varphi_F(0)&-&\varphi_N(0)\nonumber\\
&=&r_c^L[1-(\Delta\Sigma_L/\Sigma_L)\Gamma_L]J-
{\case 1/2}(\Delta\rho_L/\rho_F^L)\zeta_F(0).
\label{eq4.3}
\end{eqnarray}
Therefore, even for a zero  contact resistance, $r_c^L=0$, when the electrochemical potentials are continuous, the electrical potential $\varphi(x)$ is discontinuous. This discontinuity originates from the non-equilibrium spins, $\zeta_F(0)\neq 0$, in the contact. More exactly, this potential drop that appears as a discontinuity in $\varphi(x)$ in the framework of the quasineutral theory, develops as a drop of a continuously changing potential at the scale of about the screening length ${\it l}_{\rm sc}$.

Using Eq.~(\ref{eq1.11}), one can eliminate $Z_F(x)$ from Eqs.~(\ref{eq1.1}) and (\ref{eq1.5}) and relate the electrical field $-\nabla\varphi_F(x)$ to $\nabla\zeta_F(x)$:
\begin{equation}
\nabla\varphi_F(x)=
\case 1/2(\Delta\sigma/\sigma_F-\Delta\rho/\rho_F)\nabla\zeta_F(x)-J/\sigma_F.
\label{eq4.4}
\end{equation}
In a similar way, it follows from Eqs.~(\ref{eq1.2}), (\ref{eq1.3}), and (\ref{eq4.1}) that the electron spin polarization equals
\begin{equation}
n_F(x)\equiv n_\uparrow^F(x)-n_\downarrow^F(x)=
2e(\rho_\uparrow\rho_\downarrow/\rho_F)\zeta_F(x).
\label{eq4.5}
\end{equation}
Notice, that in the right hand sides of Eqs.~(\ref{eq1.4}) and (\ref{eq4.5}) the same factor $\rho_\uparrow\rho_\downarrow/\rho_F$ appears. This fact indicates the self-consistency of its choice. It should be emphasized that Eq.~(\ref{eq1.4}) is quite general and can be derived without invoking the quasineutrality condition. In a normal conductor $\nabla\varphi_N(x)=-J/\sigma_N$ and 
$n_N(x)=(e\rho_N/2)\zeta_N(x)$. 

Therefore, in the quasineutral regime the concentrations $n_F(x)$ and $n_N(x)$ that are measured in optical experiments can be expressed in terms of the electrochemical potentials found in Secs.~\ref{sec:FN} and \ref{sec:FNF}.

Using Eqs.~(\ref{eq4.5}) and (\ref{eq1.11a}), one can rewrite Eq.~(\ref{eq1.9}) for $\Gamma_F(x)$ in the conventional form through the concentration gradient
\begin{equation}
\Gamma_F(x)=eD_F\nabla n_F(x)/J+\Delta\sigma/\sigma_F.
\label{eq4.6}
\end{equation}
This equation is valid when the quasineutrality condition, Eq.~(\ref{eq4.1}), is fulfilled.

\subsection{Spin-e.m.f.}
\label{sec:emf}

Any kind of the non-equilibrium in an electronic system results in an electro-motive force.\cite{LL36} This general statement is completely applicable to spin non-equilibrium. A Dember type spin-e.m.f. develops in homogeneous magnetized media \cite{DP71} while a ``valve" spin-e.m.f. arises at the spin selective contacts. 
Measuring the spin-e.m.f. in metals has been used for detecting the electrical spin-injection,\cite{JS85,J93} while the absence (or a very small magnitude) of the spin-e.m.f. signal from semiconductor microstructures prompted questioning feasibility of the spin injection from ferromagnetic metals into semiconductors.\cite{Fil00} 

In this section the spin-e.m.f. in a F-N-junction is considered under the conditions when a homogeneous spin pumping of the N region produces a non-equilibrium spin magnetization. It can be produced either electrically or optically. All arguments are similar to those of Sec.~\ref{sec:FN} with the only difference that
\begin{equation}
\zeta_N(x)=\zeta_\infty^N+(\zeta_N-\zeta_\infty^N)\exp(-x/L_N),
\label{eq4.7}
\end{equation}
where $\zeta_\infty^N=\zeta_N(x\rightarrow \infty)$. Because the open circuit regime is of the main importance, the equation $j_F(0)=j_N(0)\equiv j(0)$ should be used instead of Eq.~(\ref{eq1.14}). When $J=0$, this equation and Eq.~(\ref{eq1.16}) result in the following solution for the electrochemical potentials on the both sides of the junction and for the spin current at $x=0$:
\begin{eqnarray}
\zeta_F=(r_N/r_{FN})\zeta_\infty^N&,&~~\zeta_N=[(r_c+r_F)/r_{FN}]\zeta_\infty^N,\nonumber\\
j(0)&=&\zeta_\infty^N/2r_{FN}.
\label{eq4.8}
\end{eqnarray}
Substituting this solution into Eq.~(\ref{eq4.3}) allows to find the spin-e.m.f. at the contact, $\Delta\varphi_c=\varphi_N(0)-\varphi_F(0)$:
\begin{equation}
\Delta\varphi_c=\left(r_F{{\Delta\rho}\over{\rho_F}}+r_c{{\Delta\Sigma}\over\Sigma}\right)
{{\zeta_\infty^N}\over{2r_{FN}}}.
\label{eq4.9}
\end{equation}

This spin-e.m.f. develops at the contact in the region narrow compared to $L_F$ and $L_N$. The Dember type spin-e.m.f. that develops in the quasineutral regions on the both sides of the contact where the concentrations are $x$-dependent can be found by using Eqs.~(\ref{eq2.6}) with $J=0$. This procedure results in the total spin-e.m.f. at the F-N-junction:
\begin{equation}
\Delta\varphi_{FN}=\varphi_N(\infty)-\varphi_F(-\infty)=\case 1/2 \gamma \zeta_\infty^N,
\label{eq4.10}
\end{equation}
where $\gamma$ is the spin injection coefficient of Eq.~(\ref{eq2.4}). The right hand side of Eq.~(\ref{eq4.10}) has been calculated as the difference of the electrochemical potentials, $Z_F(-\infty)-Z_N(\infty)$, and the equation for this difference is rather general. Under the quasineutrality conditions this difference in $Z$'s also equals the drop of the electrical potential across the specimen as is seen from Eq.~(\ref{eq4.2}). Indeed, the second term in Eq.~(\ref{eq4.2}) turns into zero at $x=-\infty$ because of $\zeta_F(-\infty)=0$, i.e., because of the absence of the non-equilibrium spins, while at $x=\infty$ it turns into zero (despite the spin non-equilibrium at $x=+\infty$) because of $\Delta\rho=0$ in the N conductor. 

The difference between $\Delta\varphi_c$ and $\Delta\varphi_{FN}$ comes completely from the Dember e.m.f. in the F region. The contribution to this difference from the N region equals zero because of $\Delta\sigma=\Delta\rho=0$.

Because $\zeta_\infty^N=2n_\infty^N/e\rho_N$, both $\Delta\varphi_c$ and $\Delta\varphi_{FN}$ can be also expressed in terms of $n_\infty^N=n_N(x=\infty).$ After such a transformation, Eq.~(\ref{eq4.10}) becomes identical to Eq.~(19) of 
Ref.~\onlinecite{R00}.

It is seen from Eqs.~(\ref{eq4.9}) and (\ref{eq4.10}) that the criterion for a large spin-e.m.f. is the same as the criterion for a large spin injection coefficient $\gamma$: the resistance $r_N$ should be less than at least one of the resistances $r_c$ and 
$r_F$.

\section{Impedance of a F-N-junction}
\label{sec:impedance}

The resistances of F-N- and F-N-F-junctions have been found in Secs.~\ref{sec:FN} and \ref{sec:resist} without the requirement of a local quasineutrality, Eq.~(\ref{eq4.1}). However, it is only possible for a time-independent voltage. To find a response to a time-dependent voltage, the Poisson equation or the quasineutrality condition should be employed explicitly. In this section the frequency dependence of the complex impedance ${\cal Z}(\omega)$ of a spin selective junction will be calculated. Similar to a $p-n$-junction, the reactive part of $\cal Z$ includes a geometrical capacitance $C_{\rm geom}$ and the diffusion part $C_{\rm diff}(\omega)$ controlled by the spin injection and spin relaxation. The frequency dependence of $C_{\rm diff}(\omega)$ is indicative of the spin relaxation rate. It will be found in what follows. The geometrical capacitance $C_{\rm geom}$ can be estimated as $C_{\rm geom}\approx\varepsilon/4\pi X$, where $X$ is about the contact width for a tunnel contact and about several screening lengths ${\it l}_{\rm sc}$ for a Schottky contact. Because the geometrical contribution to $\cal Z$ depends on the detailed structure of the contact it cannot be found in a general form. However, the frequency dependence of $C_{\rm geom}$ is expected to be slow, and this fact should allow to separate $C_{\rm geom}$ and $C_{\rm diff}(\omega)$ experimentally. 

When the voltage depends on the time, Eq.~(\ref{eq1.4}) should be substituted with
\begin{equation}
{1\over e}\nabla j_\uparrow(x,t)=
{{e\rho_\uparrow\rho_\downarrow}\over{\rho_\uparrow+\rho_\downarrow}}
{{\zeta_\uparrow(x,t)-\zeta_\downarrow(x,t)}\over{\tau_s^F}}
+\partial_t n_\uparrow(x,t).
\label{eq5.1}
\end{equation}
Using Eq.~(\ref{eq1.2}), one can eliminate $n_\uparrow(x,t)$ from 
Eq.~(\ref{eq5.1}). After a similar transformation of the equation for 
$j_\downarrow(x,t)$ and changing to the symmetric variables of 
Eqs.~(\ref{eq1.6}) and (\ref{eq1.7}), the following equations emerge:
\end{multicols}
\widetext 
\begin{equation}
\nabla^2\zeta_F(x,t)=\zeta_F(x,t)/L_F^2
+[(D_\uparrow+D_\downarrow)/2D_\uparrow D_\downarrow]\partial_t\zeta_F(x,t)-
[(D_\uparrow-D_\downarrow)/D_\uparrow D_\downarrow]
\partial_t[\varphi_F(x,t)+Z_F(x,t)],
\label{eq5.2}
\end{equation}
\begin{equation}
\nabla^2Z_F(x,t)=-(\Delta\sigma/2\sigma_F)\zeta_F(x,t)/L_F^2
+[(D_\uparrow+D_\downarrow)/2D_\uparrow D_\downarrow]
\partial_t[\varphi_F(x,t)+Z_F(x,t)]-
[(D_\uparrow-D_\downarrow)/4D_\uparrow D_\downarrow]
\partial_t\zeta_F(x,t).
\label{eq5.3}
\end{equation}
\begin{multicols}{2}
It has been emphasized at the end of Sec.~\ref{sec:basic} that in the dc regime the equations for the functions $\zeta(x)$ and $Z(x)$ completely separate, and the potential $\varphi(x)$ does not enter them. On the contrary, in the dynamic (ac) regime all these functions are entangled as it is seen from Eqs.~(\ref{eq5.2}) and 
(\ref{eq5.3}). To make the problem soluble analytically, it is needed to apply the quasineutrality condition from the very beginning. Eliminating $n_\uparrow(x,t)$ and $n_\downarrow(x,t)$ from the equation 
$\partial_t[n_\uparrow(x,t)+n_\downarrow(x,t)]=0$ by employing Eqs.~(\ref{eq1.2}) results in
\begin{equation}
\partial_t[\varphi_F(x,t)+Z_F(x,t)]
+{\case 1/2}(\Delta\rho/\rho_F)\partial_t\zeta_F(x,t)=0.
\label{eq5.4}
\end{equation}
Eqs.~(\ref{eq5.4}) allows to eliminate $\varphi_F(x,t)$ from Eqs.~(\ref{eq5.2}) and (\ref{eq5.3}). After some algebra, equations for $\zeta_F(x,t)$ and $Z_F(x,t)$ simplify essentially:
\begin{equation}
\nabla^2\zeta_F(x,t)=\zeta_F(x,t)/L_F^2+\partial_t\zeta(x,t)/D_F.
\label{eq5.5}
\end{equation}
\begin{equation}
\nabla^2Z_F(x,t)=-(\Delta\sigma/2\sigma_F)[\zeta_F(x,t)/L_F^2
+\partial_t\zeta(x,t)/D_F].
\label{eq5.6}
\end{equation}
Substituting Eq.~(\ref{eq5.5}) into the right hand side of Eq.~(\ref{eq5.6}) results in the equation 
$\nabla^2Z_F(x,t)=-(\Delta\sigma/2\sigma_F)\nabla^2\zeta_F(x,t)$ having an obvious first integral
\begin{equation}
\nabla Z_F(x,t)=
-(\Delta\sigma/2\sigma_F)\nabla \zeta_F(x,t)+J(t)/\sigma_F.
\label{eq5.6a}
\end{equation}
This equation can be derived directly from the conservation of the total current
 $J(t)$, Eq.~(\ref{eq1.5}). Similar equations operate in the N region.

It is an important property of Eqs.~(\ref{eq5.5}) -- (\ref{eq5.6a}) that for a harmonic signal, $\exp(-i\omega t)$, they are equivalent to Eqs.~(\ref{eq1.10}) and (\ref{eq1.11}) with the only difference that instead of $L_F$ and $L_N$ the frequency dependent diffusion lengths $L_F(\omega)$ and $L_N(\omega)$ should be used:
\begin{equation}
{1\over{L^2_F(\omega)}}={1\over{L^2_F}}(1-i\omega\tau_s^F),~
{1\over{L^2_N(\omega)}}={1\over{L^2_N}}(1-i\omega\tau_s^N).
\label{eq5.7}
\end{equation}
Therefore, the frequency dependence appears also in the effective resistances, 
$r_F(\omega)$ and $r_N(\omega)$, that are defined through $L_F(\omega)$ and 
$L_N(\omega)$ by Eq.~(\ref{eq2.3}). Through these resistances, the spin injection coefficient $\gamma(\omega)$, Eq.~(\ref{eq2.4}), and in the resistance $R$, Eq.~(\ref{eq2.8}), also acquire imaginary parts. The phase of $\gamma(\omega)$ describes the phase shift between the spin current $j(x=0, t)$ and the total current $J(t)$, and, therefore, allows to find the phase shift between the spin magnetization and $J(t)$.

The frequency dependence of ${\rm Re}\{{\cal Z}\}$ and ${\rm Im}\{{\cal Z}\}$ provides a useful tool for measuring different parameters of a F-N-junction. Some general regularities are seen directly from equations (\ref{eq5.7}). The spin relaxations times $\tau_s^F$ and $\tau_s^N$ in a ferromagnetic aligner and in a semiconductor microstructure, respectively, may differ by several orders of magnitude. Usually $\tau_s^F\ll \tau_s^N$. Therefore, two different scales, $(\tau_s^F)^{-1}$ and $(\tau_s^N)^{-1}$, should be seen in the frequency dependence of ${\cal Z}(\omega)$, and this fact shows the way for measuring $r_F$ and $r_N$. At high frequencies, $\omega\gg (\tau_s^F)^{-1},(\tau_s^N)^{-1}$,  the diffusive contribution to $\cal Z$ vanishes and ${\cal Z}\rightarrow R_{\rm eq}$.

An explicit expression for the impedance ${\cal Z}(\omega)$ can be found by plugging $r_F(\omega)$ and $r_N(\omega)$ into Eq.~(\ref{eq2.8}):
\begin{eqnarray}
 {\cal Z}_{\rm n-eq}(\omega)&&\nonumber\\
={1\over{r_{FN}(\omega)}}
&\{& r_N(\omega)\left[r_c(\Delta\Sigma/\Sigma)^2
+r_F(\omega)(\Delta\sigma/\sigma_F)^2 \right]\nonumber \\
&+&r_c r_F(\omega)\left[(\Delta\Sigma/\Sigma)-(\Delta\sigma/\sigma_F)\right]^2 \},
\label{eqZ}
\end{eqnarray}
in agreement with Ref.~\onlinecite{R02}. Here
\begin{equation}
r_F(\omega)=L_F(\omega)/\sigma_F,~~r_N(\omega)=L_N(\omega)/\sigma_N .
\label{eqr}
\end{equation}

It is instructive to consider the low- and high-frequency regimes in more detail. In the low-frequency regime, $\omega\ll (\tau_s^F)^{-1},(\tau_s^N)^{-1}$, the real part of the impedance ${\rm Re}\{{\cal Z}(\omega)\}\approx R$. The expansion of
${\rm Im}\{{\cal Z}(\omega)\}$ in $\omega$ begins with a positive linear in $\omega$ term, and the sign of ${\rm Im}\{\cal Z(\omega)\}$ suggests that 
${\rm Im}\{{\cal Z}^{-1}(\omega)\}$ can be considered as the complex conductivity $-i\omega C_{\rm diff}$\cite{LL93} of a capacitor $C_{\rm diff}$ connected in parallel to the resistor $R$:
\begin{eqnarray}
C_{\rm diff}&=&\biggl\{\tau_s^N r_N\left(r_c{{\Delta\Sigma}\over\Sigma}
+r_F{{\Delta\sigma}\over{\sigma_F}}\right)^2 \nonumber\\
&+&\tau_s^Fr_F\left[r_c{{\Delta\Sigma}\over\Sigma}
-(r_c+r_N){{\Delta\sigma}\over{\sigma_F}}
\right]^2
\biggr\}/2R^2r_{FN}^2.
\label{eq5.8}
\end{eqnarray}
It is seen from Eq.~(\ref{eq5.8}) that $C_{\rm diff}>0$ for arbitrary values of the parameters.

The diffusion capacitance $C_{\rm diff}$ in its essence is similar to the diffusion capacitance of a $p-n$-junction.\cite{Sh49} However, its dependence on the relaxation time $\tau$ is rather different for $p-n$- and F-N-junctions. The square root dependence, $C_{\rm diff}\propto \tau^{1/2}_s$, is typical of $p-n$-junctions. The diffusion capacitance of F-N-junctions follows this law only when $r_c\ll r_F,~r_N$, i.e., when the spin injection from a metal into a semiconductor is strongly suppressed. In the opposite regime $r_c\gg r_F,~r_N$, that is of major importance for the spin injecting devices, it follows from Eq.~(\ref{eq5.8}) that $C_{\rm diff}\propto \tau_s^{3/2}$. Depending on the relative magnitude of $r_F$ and $r_N$, different combinations of $\tau_s^F$ and $\tau_s^N$ can appear in $C_{\rm diff}$. Therefore, a large $\tau_s^N$ typical of semiconductor microstructures\cite{KA98,Hag98} should enlarge $C_{\rm diff}$ considerably.

In the opposite limit of a high frequency, when
$\omega\gg (\tau_s^F)^{-1},(\tau_s^N)^{-1}$, the resistances 
$r_F$ and $r_N$ are small and one can expand ${\cal Z}_{\rm diff}(\omega)$ in $r_F/r_c,~ r_N/r_c\ll 1$. Because in this limit $L_F(\omega)\approx(1+i)L_F/\sqrt{2\omega\tau_s^F}$ and similarly for $L_N(\omega)$, the real and imaginary parts of 
${\cal Z}_{\rm diff}(\omega)$ are nearly equal:
\begin{equation}
R_{\rm n-eq}(\omega)={1\over{\sqrt{2\omega}}}
\left[{{r_N}\over{\sqrt{\tau_N}}}
\left({{\Delta\Sigma}\over\Sigma}\right)^2
+{{r_F}\over{\sqrt{\tau_F}}}
\left({{\Delta\Sigma}\over\Sigma}-{{\Delta\sigma}\over{\sigma_F}}\right)^2
\right]
\label{eq5.9}
\end{equation}
\begin{equation}
C_{\rm diff}(\omega)=R_{\rm n-eq}(\omega)/\omega R_{\rm eq}^2.
\label{eq5.10}
\end{equation}
Therefore, $R_{\rm n-eq}(\omega)>0$ and $C_{\rm diff}(\omega)>0$ also in this limit. Because both $R_{\rm n-eq}(\omega)$ and $C_{\rm diff}(\omega)$ decay with $\omega$, only $R_{\rm eq}$ and $C_{\rm geom}$ survive when $\omega\rightarrow\infty$.

The impedance ${\cal Z}(\omega)$ of a F-N-junction has been found by combining Eqs.~(\ref{eq5.7}) for the spin-diffusion lengths with Eq.~(\ref{eq2.8}) for the dc resistance. In a similar way, one can find the impedance of a F-N-F-junction using Eq.~(\ref{eq3.18}) or Eqs.~(\ref{A1}) and (\ref{A2}) for its dc resistance. Because these equations are rather long, such a procedure might prove to be useful mostly in some limit cases or for the numerical processing the experimental data. 

\section{Spin non-conserving F-N-junction}
\label{sec:non-cons}

The F-N-junctions considered in Sec.~\ref{sec:FN} and everywhere above were spin selective, $\Sigma_{\uparrow}\neq\Sigma_{\downarrow}$, however, the strict spin conservation condition, $j_F(0)=j_N(0)$, has been imposed on them. Meanwhile, the spin orbit interaction and the inhomogeneity of the exchange field should result in the spin dynamics and spin relaxation inside the contacts. This processes can be taken into account by generalizing 
Eq.~(\ref{eq1.15}) as
\begin{eqnarray}
j_{\uparrow}^F(0)&=&\Sigma_{\uparrow\uparrow}(\zeta^N_{\uparrow}-\zeta^F_{\uparrow})
+\Sigma_{\uparrow\downarrow}(\zeta^N_{\downarrow}-\zeta^F_{\uparrow}),\nonumber\\
j_{\downarrow}^F(0)&=&\Sigma_{\downarrow\uparrow}(\zeta^N_{\uparrow}-\zeta^F_{\downarrow})
+\Sigma_{\downarrow\downarrow}(\zeta^N_{\downarrow}-\zeta^F_{\downarrow}).
\label{eq6.1}
\end{eqnarray}
Here the conductivities $\Sigma_{\alpha\beta}$ describe the transfer of an electron from the $\alpha$ spin state in the ferromagnet into the $\beta$ spin state in the normal conductor. It is the only restriction imposed on $\Sigma_{\alpha\beta}$ that they are positive, $\Sigma_{\alpha\beta}>0$. The Onsager type relations are not applicable because the time inversion symmetry is brocken in the F region. Similar equations can be written for $j_{\uparrow}^N(0)$ and $j_{\downarrow}^N(0)$. These equations conserve the total current $J$. 

It is convenient to change to the symmetric variables of Eqs.~(\ref{eq1.6}) and (\ref{eq1.7}). Then the boundary conditions take a form
\begin{eqnarray}
&j&_F(0)-j_N(0)=-2\Sigma_a(Z_N-Z_F)-\Sigma_s(\zeta_F+\zeta_N),\nonumber\\
&j&_F(0)+j_N(0)=2\Delta\Sigma (Z_N-Z_F)-\Sigma_d(\zeta_F-\zeta_N),\nonumber\\
&J&=\Sigma(Z_N-Z_F)-{\case 1/2}\Delta\Sigma(\zeta_F-\zeta_N)
+{\case 1/2}\Sigma_a(\zeta_F+\zeta_N).
\label{eq6.2}
\end{eqnarray}
In equations (\ref{eq6.1}) and (\ref{eq6.2}) all potentials  $\zeta$ and $Z$ with the subscripts (or superscripts) $F$ and $N$ are taken at the points $x=0-$ and $x=0+$, respectively. The contact conductivities $\Sigma$ with different indices appearing in 
Eq.~(\ref{eq6.2}) are defined as:
\begin{eqnarray}
\Sigma_d=\Sigma_{\uparrow\uparrow}&+&\Sigma_{\downarrow\downarrow},~~
\Sigma_s=\Sigma_{\downarrow\uparrow}+\Sigma_{\uparrow\downarrow},~~
\Sigma=\Sigma_d+\Sigma_s,\nonumber\\
\Delta\Sigma&=&\Sigma_{\uparrow\uparrow}-\Sigma_{\downarrow\downarrow},~~
\Sigma_a=\Sigma_{\downarrow\uparrow}-\Sigma_{\uparrow\downarrow}.
\label{eq6.3}
\end{eqnarray}
Here $\Sigma_d$ and $\Delta\Sigma$ are the symmetric and antisymmetric combinations of the diagonal elements of the matrix $\Sigma_{\alpha\beta}$, respectively, and 
$\Sigma_s$ and $\Sigma_a$ are the similar combinations of its non-diagonal components.

It has been emphasized in Sec.~\ref{sec:basic} that in the dc regime the equations for $\zeta_{F(N)}$ and $j_{F(N)}$ separate from the equations for $Z_{F(N)}$ when the contacts are spin-conserving and, therefore, these equations can be solved independently from the equations for $Z_{F(N)}$. It is not the case for the spin non-conserving contacts. It is seen from Eq.~(\ref{eq6.2}) that the difference $Z_N-Z_F$ cancels out from the equations for spin-currents $j_F(0)$ and $j_N(0)$ (and, therefore, the equations decouple) when and only when
\begin{equation}
\Delta\Sigma =0,~~\Sigma_a=0.
\label{eq6.4}
\end{equation}
The contacts obeying these conditions will be termed spin relaxors in what follows because they reduce the spin polarized current originating from the bulk polarization of the ferromagnet, $\Delta\sigma\neq 0$. This property will be discussed in more detail below in the context of Eqs.~(\ref{eq6.9}) and (\ref{eq6.10}) for the spin injection coefficients.

It is convenient to define the spin injection coefficients $\gamma_F=j_F(0)/J$ and 
$\gamma_N=j_N(0)/J$ on both sides of the contact. Because the contact is spin non-conserving, $\gamma_F\neq\gamma_N$, the number of the unknowns is doubled compared to the only unknown $\gamma$ for a spin-conserving contact, cf. 
Sec.~\ref{sec:FN}. Therefore, two equations are needed to find them. Using the third equation (\ref{eq6.2}), one can eliminate $Z_N-Z_F$ from the first two equations. Afterwards, one applies Eq.~(\ref{eq2.2}) to eliminate $\zeta_F$ and $\zeta_N$. Finally, the system of two equations of the $\gamma$-technique for two variables
\begin{equation}
\gamma_{+}=\gamma_F +\gamma_N,~~\gamma_{-}=\gamma_F -\gamma_N
\label{eq6.5}
\end{equation}
emerges. It has the form
\begin{eqnarray}
(r_FC_+-r_NC_-&-&1)\gamma_-+(r_FC_++r_NC_-)\gamma_+\nonumber\\
&=&2\Sigma_a/\Sigma+2r_FC_+(\Delta\sigma/\sigma_F)\nonumber\\
(r_F{\tilde C}_++r_N{\tilde C}_-)&\gamma_-&+(r_F{\tilde C}_+-r_N{\tilde C}_--1)\gamma_+\nonumber\\
&=&-2\Delta\Sigma/\Sigma+2r_F{\tilde C}_+(\Delta\sigma/\sigma_F),
\label{eq6.6}
\end{eqnarray}
where
\begin{eqnarray}
C_{\pm}&=&-\Sigma_a\Delta\Sigma/\Sigma\pm(\Sigma_a^2/\Sigma-\Sigma_s),\nonumber\\
{\tilde C}_{\pm}&=&-\Sigma_a\Delta\Sigma/\Sigma\pm(\Delta\Sigma^2/\Sigma-\Sigma_d).
\label{eq6.7}
\end{eqnarray}
The determinant of this system ${\cal D}^{\prime}$ is obviously positive
\begin{eqnarray}
{\cal D}^{\prime}=1&+&4r_F(\Sigma_{\uparrow\uparrow}+\Sigma_{\uparrow\downarrow})
(\Sigma_{\downarrow\downarrow}+\Sigma_{\downarrow\uparrow})/\Sigma\nonumber\\
&+&4r_N(\Sigma_{\uparrow\uparrow}+\Sigma_{\downarrow\uparrow})
(\Sigma_{\downarrow\downarrow}+\Sigma_{\uparrow\downarrow})/\Sigma\nonumber\\
&+&16r_Fr_N(\Sigma_s\Sigma_{\uparrow\uparrow}\Sigma_{\downarrow\downarrow}
+\Sigma_d\Sigma_{\uparrow\downarrow}\Sigma_{\downarrow\uparrow})/\Sigma.
\label{eq6.8}
\end{eqnarray}

It is instructive to consider the spin injection coefficients $\gamma_-$ and 
$\gamma_N$:
\begin{eqnarray}
{\cal D}^{\prime}\gamma_-&=&-2\Sigma_a/\Sigma+4r_N(\Sigma_{\uparrow\uparrow}
\Sigma_{\uparrow\downarrow}-\Sigma_{\downarrow\downarrow}\Sigma_{\downarrow\uparrow})/\Sigma \nonumber\\
&+&4r_F[(\Sigma_{\downarrow\downarrow}\Sigma_{\uparrow\downarrow}-\Sigma_{\uparrow\uparrow}\Sigma_{\downarrow\uparrow})\nonumber\\
&+&(\Sigma_{\uparrow\uparrow}\Sigma_{\downarrow\uparrow}+\Sigma_{\downarrow\downarrow}\Sigma_{\uparrow\downarrow}+2\Sigma_{\uparrow\downarrow}\Sigma_{\downarrow\uparrow})(\Delta\sigma/\sigma_F)]/\Sigma\nonumber\\
+&16&r_Fr_N(\Sigma_d\Sigma_{\uparrow\downarrow}\Sigma_{\downarrow\uparrow}+
\Sigma_s\Sigma_{\uparrow\uparrow}\Sigma_{\downarrow\downarrow})(\Delta\sigma/\sigma_F)/\Sigma,
\label{eq6.9}
\end{eqnarray}
\begin{eqnarray}
{\cal D}^{\prime}\gamma_N=(\Delta\Sigma&+&\Sigma_a)/\Sigma
+4r_F[(\Sigma_{\uparrow\uparrow}\Sigma_{\downarrow\uparrow}-\Sigma_{\downarrow\downarrow}\Sigma_{\uparrow\downarrow})\nonumber\\
&+&(\Sigma_{\uparrow\uparrow}\Sigma_{\downarrow\downarrow}-\Sigma_{\downarrow\uparrow}\Sigma_{\uparrow\downarrow})(\Delta\sigma/\sigma_F)]/\Sigma.
\label{eq6.10}
\end{eqnarray}
The coefficient $\gamma_-$ describes the change in the spin polarized current between the left and the right hand sides of the contact. It is seen from Eq.~(\ref{eq6.9}) that the sign of $\gamma_-$ depends on the relative magnitude of several competing parameters. This fact indicates that inside the contact the polarization of the current can both increase or decrease. 

Eq.~(\ref{eq6.4}) is the only general condition that can fix the sign of $\gamma_-$  irrespective to the relative values of the different coefficients 
$\Sigma_{\alpha\beta}$. If it is satisfied,  then
\begin{eqnarray}
&{\cal D}^{\prime}&\gamma_-=2r_F\Sigma_s(1+2r_N\Sigma_d)(\Delta\sigma/\sigma_F),\nonumber\\
&{\cal D}^{\prime}&=1+(r_F+r_N)\Sigma+4r_Fr_N\Sigma_d\Sigma_s,
\label{eq6.11}
\end{eqnarray}
hence, $\gamma_-$ has the sign of $\Delta\sigma$ (that determines the polarization of the current in the bulk). In this case $\gamma_-$ describes the loss in the polarization inside the contact that increases monotonically with $\Sigma_s$. For this reason, the term ``spin relaxor" has been proposed above for such a spin non-conserving contact. However, considering $\gamma_N$ shows that the real situation is more involved and, therefore, more interesting. Indeed, in the same limit the spin injection coefficient into the N region, $\gamma_N$, equals
\begin{equation}
{\cal D}^{\prime}\gamma_N=r_F(\Sigma_d-\Sigma_s)(\Delta\sigma/\sigma_F).
\label{eq6.12}
\end{equation}
With increasing $\Sigma_s$, the spin injection first decreases as is expected for the spin relaxation regime in the contact. However, when $\Sigma_s$ becomes large enough, $\Sigma_s>\Sigma_d$, the coefficient $\gamma_N$ changes its sign. It happens because $\gamma_-$ increases with $\Sigma_s$ faster than $\gamma_F$, and finally $\gamma_F$ and $\gamma_N$ acquire opposite signs. Therefore, the phenomenological theory predicts a counterintuitive behavior of the spin injection across spin non-conserving contacts.

The general equation for the junction resistance is cumbersome and, therefore, not very instructive. However, some interesting regularities can be understood taking  the spin relaxors as an example because the equation for $R$ simplifies essentially for them. With $R_{\rm eq}=\Sigma_d^{-1}$, one finds
\begin{equation}
R_{\rm n-eq}=-{{\Sigma_s}\over{\Sigma\Sigma_d}}
+{{r_F(1+r_N\Sigma)}\over{{\cal D}^\prime}}
\left({{\Delta\sigma}\over{\sigma_F}}\right)^2.
\label{eq6.13}
\end{equation}
It has been shown in Secs.~\ref{sec:FN} and \ref{sec:FNF} that for spin-conserving
contacts $R_{\rm n-eq}$ is always positive, $R_{\rm n-eq}>0$. It is not the case for Eq.~(\ref{eq6.13}). The first term describing the spin non-conservation in the contact is always negative, while the second term, related mostly to the spin relaxation in the bulk, is positive. Therefore, the sign of $R_{\rm n-eq}$ is controlled by the balance of these two terms. 

It follows from Eq.~(\ref{eq6.10}) that even when $\vert\Delta\sigma/\sigma_F\vert\ll 1$, i.e., when the spin polarization of the bulk current can be neglected (what is compatible with a large bulk magnetization), spin injection is still possible  due to $\Delta\Sigma\neq 0$ and (or) $\Sigma_a\neq 0$. Recent experiments on spin injection from metals into semiconductors through resistive contacts\cite{Plo01,HJ01,Jon01,Saf01,Weiss02,Jap02,Samarth02} indicate that the contacts play a critical role in spin injection. However, a little is knows about the microscopic mechanisms of spin injection through these contacts, spin relaxation in them, and the magnetization pattern in the contact region. Experimental data on semiconductor systems show that even when no special precautions are made to suppress the spin relaxation (and spin dynamics) in the contacts, optically induced spin coherence is largely preserved when spins cross interfaces\cite{Mal00}, and efficient electrical spin injection can be achieved in non-lattice-matched heterostructures.\cite{Jon00}  Nevertheless, the general symmetry arguments suggest that an interfacial inversion asymmetry should be present in all F-N-junctions.\cite{RS91IP97} The experimental observation of the effect of this asymmetry on the optical spectra and the spin relaxation has been already reported.\cite{KVOles} In metals, the unexpected correlation between the bulk magnetization and the spin injection into paramagnetic metals\cite{MPT76} as well as the opposite sign of the bulk magnetization and the injection\cite{WG00} were observed. Inversion of the sign of the tunnel magnetoresistance depending of the type of the tunnel contact, that can be ascribed to the $s-d$ hybridization,\cite{HA73} has been also reported.\cite{Teresa99} Even a simplified kinematic model of the tunneling of Bloch electrons results in a number of new features,\cite{M01} and recent numerical studies proved the importance of the consistent employing the realistic Bloch functions.\cite{Bloch} The spin dynamics in the contact area deserves a detailed study. According to the Stoner criterion, the Zeeman frequency in the exchange field is large, about $E_F/\hbar$, $E_F$ being the Fermi energy. Therefore, a considerable change in the electron spin direction can be accumulated during the passage through a narrow boundary layer. It is important to establish what kind of microscopic environment can result  in different regimes that the above macroscopic theory predicts. 

It is still unclear which kind of resistive contacts, tunnel contacts or Schottky contacts (promoted by Grundler\cite{G01}), are most promising for achieving efficient spin injection. First breakthrough in spin injection from metals into semiconductors has been achieved with Schottky contacts, \cite{Plo01} however, some limitations inherent in them have been critically discussed.\cite{SM02} There are impressing experimental data on the profound effect of the atomic arrangement at the ferromagnet/semiconductor interface on spin injection,\cite{Ploog02} and a first-principles theoretical study  has revealed interesting pattern of the Fe film grow on GaAs.\cite{ELS02}

\section{Conclusions}
\label{sec:concl}

Resistive contacts are the key for efficient spin-injection from ferromagnetic metals into semiconductors.\cite{R00} However, including the contact resistances into the diffusion theory of spin injection makes the equations cumbersome. It becomes difficult to derive explicit formulae for the injection coefficients and the contact resistances and even to establish the basic qualitative regularities. 
 
In this paper, the $\gamma$-technique based on deriving a system of self-consistent equations for spin-injection coefficients $\gamma$'s (or $\Gamma$'s) is developed. It permits one to simplify significantly the procedure of solving the system of the diffusion equations for the electrochemical potentials $\zeta(x)$. In this technique the values of $\zeta$'s at all interfaces are expressed through $\gamma$'s and the self-consistency equations for $\gamma$'s are derived. They are concise when written in appropriate notations. The system of two equations, Eq.~(\ref{eq3.4}), for the spin-injection coefficients $\Gamma_L$ and $\Gamma_R$ through the left and the right contact of a F-N-F-junction is an example. The resistance of the junction can be found in the same way. Eq.~(\ref{eq3.18}) for the resistance of a F-N-F-junction $\cal R$ is valid even for non-symmetric junctions. The technique allows to isolate in a natural way the part of $\cal R$ describing the spin-valve effect. Both these quantities can be measured experimentally.

Because all calculations are well organized in the $\gamma$-technique, it not only permits one to solve problems that were out of reach before but also increases essentially the crediability of the results. I mention in this context that Eqs.~(\ref{eq3.19a}) and (\ref{eq3.20}) for symmetric F-N-F-junctions differ from the results that can be found in the literature.

The application of $\gamma$-technique allowed to establish some general properties of spin-injecting junctions. E.g., it was shown that for spin-conserving contacts the non-equilibrium part of the resistance ${\cal R}_{\rm n-eq}$ is always positive, ${\cal R}_{\rm n-eq}>0$. However, spin non-conservation in the contacts reduces ${\cal R}_{\rm n-eq}$ and it can even change its sign. 

In the present paper the $\gamma$-technique was applied to the one-dimensional electron flow when all solutions are exact. However, I anticipate its application  also to more realistic geometries with several contacts under the conditions when some reasonable assumptions may allow deriving a system of approximate equations for $\gamma$'s.

It is important to get understanding how long the basic physical conclusions of this paper about the role of resistive contacts survive when the restrictions of the diffusion regime are relaxed. It has been argued that a tunnel barrier enhances spin emission even in a ballistic regime when some conditions are met.\cite{HSNT01} A recent Boltzmann theory proves that a large ratio of the contact resistance to the Sharvin resistance of the ballistic region is the basic criterion for efficient spin injection.\cite{KR02}

\section*{Acknowledgment}

Support from DARPA/SPINS by the Office of Naval Research Grant N000140010819 is gratefully acknowledged.

\end{multicols} 

\appendix\section*{}
\label{sec:app}

\begin{multicols}{2}

By an elementary but lengthy algebraic transformation, the non-equilibrium resistance ${\cal R}_{\rm n-eq}$ of Eq.~(\ref{eq3.18}) can be rewritten in the form
\begin{equation}
{\cal R}_{\rm n-eq}={\cal N}/{\cal D}\sinh(d/L_N),
\label{A1}
\end{equation}
where $\cal D$ is defined by Eq.~(\ref{eq3.7}) and $\cal N$ equals

\end{multicols}
\widetext 
\begin{eqnarray}
{\cal N}&=&\left\{\left[r_N^2r_F^L(\Delta\sigma_L/\sigma_F^L)^2
+r_N^2r_c^L(\Delta\Sigma_L/\Sigma_L)^2
+r_F^Lr_c^L(r_F^R+r_c^R)(\Delta\sigma_L/\sigma_F^L-\Delta\Sigma_L/\Sigma_L)^2\right]
+L\Leftrightarrow R\right\}\sinh(d/L_N) \nonumber \\
&+&\left\{\left[r_Nr_F^Lr_c^L(\Delta\sigma_L/\sigma_F^L-\Delta\Sigma_L/\Sigma_L)^2\right]
+L\Leftrightarrow R\right\}\cosh(d/L_N) \nonumber \\
&+&\left\{(r_F^R+r_c^R)r_N\left[r_F^L(\Delta\sigma_L/\sigma_F^L)^2
+r_c^L(\Delta\Sigma_L/\Sigma_L)^2\right]+L\Leftrightarrow R\right\}[\cosh(d/L_N)-1]
\nonumber \\
&+&r_N\left[r_F^Lr_F^R(\Delta\sigma_L/\sigma_F^L-\Delta\sigma_R/\sigma_F^R)^2
+r_c^Lr_c^R(\Delta\Sigma_L/\Sigma_L-\Delta\Sigma_R/\Sigma_R)^2\right]\nonumber \\
&+&r_N[r_F^Lr_c^R(\Delta\sigma_L/\sigma_F^L-\Delta\Sigma_R/\Sigma_R)^2
+L\Leftrightarrow R].
\label{A2}
\end{eqnarray}
The symbol $L\Leftrightarrow R$ stands everywhere for an expression that differs from the preceeding term by the interchange of the indeces $L$ and $R$. All terms in Eq.~(\ref{A2}) are obviously positive and ${\cal D}>0$, hence, ${\cal R}_{\rm n-eq}>0$.

\newpage
\begin{multicols}{2}

\end{multicols} 

\begin{references}
\bibitem[*]{Rashba*} Also at the Department of Physics, MIT, Cambridge, Massachusetts 02139. \\
E-mail: erashba@mailaps.org.
\bibitem{Wolf} S. A. Wolf, D. D. Awschalom, R. A. Buhrman, J. M. Daughton, 
S. von Molnar, M. L. Roukes, A. Y. Chtchelkanova, and D. M. Treger,
Science {\bf 294}, 1488 (2001).
\bibitem{DSFHZ} S. Das Sarma, J. Fabian, X. Hu, and I. \v{Z}uti\'{c},
Solid State Commun. {\bf 119}, 207 (2001).
\bibitem{DD90} S. Datta and B. Das, Appl. Phys. Lett. {\bf 56},665 1990). Different modifications of the device have been proposed recently.
\bibitem{R60} E. I. Rashba, Fiz. Tverd. Tela. {\bf 2}, 1224 (1960) [Sov. Phys. - Solid State {\bf 2}, 1109 (1960).
\bibitem{GMR} M. N. Baibich, J. M. Broto, A. Fert, F. Nguyen Van Dau, F. Petroff, P. Etienne, G. Creuzet, A. Friederich, and J. Chazelas, Phys. Rev. Lett. {\bf 61}, 2472 (1988).
\bibitem{A76} A. G. Aronov, Pis'ma Zh. Eksp. Teor. Fiz. {\bf 24}, 37 (1976) [Sov. Phys. - JETP Lett. {\bf 24}, 32 (1976)].
\bibitem{JS8788} M. Johnson and R. H. Silsbee, Phys. Rev. {\bf 35}, 4959 (1987) and {\bf 37}, 5312 (1988). 
\bibitem{JS85}  M. Johnson and R. H. Silsbee, Phys. Rev. Lett. {\bf 55}, 1790 (1985).
\bibitem{vS87} P. C. van Son, H. Kempen, and P. Wyder, Phys. Rev. Lett. {\bf 58}, 227 (1987).
\bibitem{VF93} T. Valet and A. Fert, Phys. Rev. B {\bf 48}, 7099 (1993).
\bibitem{AP76} A. G. Aronov and G. E. Pikus, Fiz. Tekh. Poluprovodn. {\bf 10}, 1177 (1976) [Sov. Phys. Semicond. {\bf 10}, 698 (1976)].
\bibitem{Ham99} P. R. Hammar, B. R. Bennett, M. J. Yang, and M. Johnson, 
Phys. Rev. Lett. {\bf 83}, 203 (1999).
\bibitem{Gard99} S. Gardelis, C. J. Smith, C. H. W. Barnes, E. H. Linfield, 
and D. A. Ritchie, Phys. Rev. B {\bf 60}, 7764 (1999). 
\bibitem{dispute} F. G. Monson, H. X. Tang, and M. L. Roukes, Phys. Rev. Lett.
{\bf 84}, 5022 (2000); B. J. van Wees, {\it ibid.}, p. 5023; P. R. Hammar, B. R. Bennett, M. J. Yang, and M. Johnson, {\it ibid.}, p. 5024.
\bibitem{Fil00} A. T. Filip, B. H. Hoving, F. J. Jedema, B. J. van Wees, B. Dutta, and S. Borghs, Phys. Rev. B {\bf 62}, 9996 (2000). 
\bibitem{Sch00}G. Schmidt, D. Ferrand, L. W. Molenkamp, A. T. Filip, B. J. van Wees, Phys. Rev. B {\bf 62}, R4790 (2000).
\bibitem{HZ97} S. Hershfield and H. L. Zhao, Phys. Rev. B {\bf 56}, 3296 (1997). I am grateful to Dr. J. Fabian and Dr. I. \v{Z}uti\'{c} for bringing this paper to my attention.
\bibitem{Oes99} M. Oestereich, J. H\"ubner, D. H\"agele, P. J. Clar, W. Heimbrodt, W. W. R\"uhle, D. E. Ashenford, and B. Lunn, Appl. Phys. Lett. {\bf 74}, 125 (1999).
\bibitem{semimagn} R. Fiederling, M. Keim, G. Reuscher, W. Ossau, G. Schmidt, A. Waag, and L. W. Molenkamp, Nature, {\bf 402}, 787 (1999); Y. Ohno, D. K. Young, B. Beschoten, F. Matsukara, H. Ohno, and D. Awschalom, {\it ibid.} p. 790. 
\bibitem{Jon00} B. T. Jonker, Y. D. Park, B. R. Bennett, H. D. Cheong, G. Kioseoglou, and A. Petrou, Phys. Rev. B {\bf 62}, 8180 (2000).
\bibitem{ZFDS02} I. \v{Z}uti\'{c}, J. Fabian, and S. Das Sarma,
Phys. Rev. Lett. {\bf 88}, 066603 (2002). 
\bibitem{Alv95} S. F. Alvorado, Phys. Rev. Lett. {\bf 75}, 513 (1995).
\bibitem{Oh98} H. Ohno, Science {\bf 281}, 951 (1998).
\bibitem{Mon98} D. J. Monsma, R. Fluttrs, and J. C. Lodder, Science {\bf 281}, 407 (1998).
\bibitem{R00} E. I. Rashba, Phys. Rev. B {\bf 62}, R16267 (2000).
\bibitem{Plo01} H. J. Zhu, M. Ramsteiner, H. Kostial, M. Wassermeier, 
H.-P. Sch\"onherr, and K. H. Ploog, Phys. Rev. Lett.  {\bf 87}, 016601 (2001).
\bibitem{HJ01} P. R. Hammar and M. Johnson, Appl. Phys. Lett. {\bf 79}, 2591 (2001) and Phys. Rev. Lett. {\bf 88}, 066806 (2002).
\bibitem{Jon01} A. T. Hanbicki, B. T. Jonker, G. Itskos, G. Kioseoglou, and A. Petrou, Appl. Phys. Lett. {\bf 80}, 1240 (2002).
\bibitem{Saf01} V. F. Motsnyi, V. I. Safarov, J. De Boeck, J. Das, W. Van Roy, E. Goovaerts, G. Borghs, Appl. Phys. Lett. {\bf 81}, 265 (2002).
\bibitem{Weiss02} S. Kreuzer, J. Moser, W. Wegscheider, D. Weiss, M. Bichler, and D. Schuh, Appl. Phys. Lett. {\bf 80}, 4582 (2002).
\bibitem{Jap02} C.-M. Hu, J. Nitta, A. Jensen, J. B. Hansen, H. Takayanagi, T. Matsuyama, D. Heitmann, and U. Merkt, J. Appl. Phys. {\bf 91}, 7251 (2002).
\bibitem{Samarth02} S. H. Chun, S. J. Potashnik, K. C. Ku, P. Schiffer, and N. Samarth, cond-mat/0207178.
\bibitem{Dutch02} F. J. Jedema, H. B. Heersche, A. T. Filip, J. J. A. Baselmans, and van Wees, Nature {\bf 416}, 713 (2002).
\bibitem{BDF02} A. Bournel, P. Dollfus, and P. Hesto, J. Magn. Magn. Mat. {\bf 240}, 217 (2002).
\bibitem{FJ01} A. Fert and H. Jaffr\`es, Phys. Rev. B {\bf 64}, 184420 (2001).
\bibitem{SS01} D. L. Smith and R. N. Silver, Phys. Rev. B {\bf 64}, 045323 (2001).
\bibitem{R02} E. I. Rashba, Appl. Phys. Lett. {\bf 80}, 2329 (2002).
\bibitem{JedW} (a) F. J. Jedema, A. T. Filip, and B. J. van Wees, 
Nature {\bf 410}, 345 (2001); (b) F. J. Jedema, M. S. Nijboer, A. T. Filip, and B. J. van Wees, J. Supercond. {\bf 15}, 27 (2002) and references therein.
\bibitem{MZ84} {\it Optical Orientation}, ed. by F. Meier and B. P. Zakharchenya (North-Holland, Amsterdam 1984)
\bibitem{KA98} J. M. Kikkawa and D. D. Awschalom, Phys. Rev. Lett. {\bf 80}, 4313 (1998).
\bibitem{ball} T. Matsuyama, C.-M. Hu, D. Grundler, G. Meier, and U. Merkt,
Phys. Rev. B {\bf 65}, 155322 (2002).
\bibitem{Rus} E. L. Ivchenko and G. E. Pikus, Pis'ma Zh. Eksp. Teor. Fiz. {\bf 27}, 640 (1978) [Sov. Phys. - JETP Lett. {\bf 27}, 604 (1978)]; V. M. Edelstein, Solid State Commun. {\bf 73}, 233 (1990).
\bibitem{West} M. Johnson, Phys. Rev. B {\bf 58}, 9635 (1998); R. H. Silsbee, 
Phys. Rev. B {\bf 63}, 155305 (2001).
\bibitem{Sch01} G. Schmidt, G. Richter, P. Grabs, C. Gould, D. Ferrand, and L. W. Molenkamp, Phys. Rev. Lett. {\bf 87}, 227203 (2001).
\bibitem{relax} For a detailed discussion of the spin relaxation term see 
Ref.~\onlinecite{HZ97}.
\bibitem{Sh49} W. Shockley, Bell Syst. Tech. J. {\bf 28}, 435 (1949).
\bibitem{T} This estimate for $T$ follows from the following arguments. Continuity of the electrochemical potential, $\zeta$, at a contact is based on the assumption of a very fast electron exchange across it, the exchange rate being controlled by the electron tunneling rate across the contact. Discontinuity of $\zeta$ at the contact can be neglected only when it is small compared to the drop of $\zeta$ in the quasineutral regions around it, hence, $\Sigma^{-1}\ll r$, where $\Sigma$ is defined by Eq.~(\ref{eq1.15}) and $r$ by Eq.~(\ref{eq2.3}). A standard kinetic estimate relates $\Sigma$ to $T$ as $\Sigma\sim e^2v\rho T$, while $r\sim L/e^2\rho D$, where $v$ is a typical electron velocity. Taking into account that $L\sim (D\tau_s)^{1/2}$ and $D\sim v^2\tau_p$, one immediately comes to the criterion $T\gg(\tau_p/\tau_s)^{1/2}$.
\bibitem{PM01} W. E. Pickett and J. Moodera, Phys. Today, May 2001, p. 39.
\bibitem{Tal} V. Dediu, M. Murgia, F. C. Matacotta, C. Taliani, and S. Barbanera, Solid State Commun. {\bf 122}, 181 (2002).
\bibitem{LaB01} Increasing $r_F$ by choosing a special crystallographic orientation of a narrow STM tip has been achieved by V. P. LaBella, D. W. Bullock, Z. Ding, C. Emery, A. Venkatesan, W. F. Oliver, G. J. Salamo, P. M. Thibado, and M. Mortazavi, Science {\bf 292}, 1518 (2001). This idea bears some similarity with the crystallographic matching of the electrodes proposed by G. Kirczenow, Phys. Rev. B {\bf 63}, 054422 (2001). 
\bibitem{resist} Notice that the ``equilibrium" resistance $\Sigma^{-1}$ differs from the ``effective" contact resistance $r_c$ appearing in the most of the equations.
\bibitem{Th81} D. J. Thouless, Phys. Rev. Lett. {\bf 47}, 972 (1981).
\bibitem{KRnon} B. Korenblum and E. I. Rashba, Phys. Rev. Lett., {\bf 89}, 096803 (2002).
\bibitem{Hag98} D. H\"agele, M. Oestreich, W. W. R\"uhle, N. Nestle, and K. Eberl,
Appl. Phys. Lett. {\bf 73}, 1580 (1998).
\bibitem{LL36} L. D. Landau and E. M. Lifshitz, Phys. Zs. Sowjet. {\bf 9}, 477 (1936).
\bibitem{DP71} M. I. D'yakonov and V. I. Perel', ZhETF Pis. Red. {\bf 13}, 206 (1971) [JETP Lett. {\bf 13}, 144 (1971)].
\bibitem{J93} M. Johnson, Phys. Rev. Lett. {\bf 70}, 2142 (1993).
\bibitem{LL93} L. D. Landau and E. M. Lifshitz, Electrodynamics of Continuous Media 
(Pergamon, Oxford-NY 1993).
\bibitem{Mal00} I. Malajovich, D. D. Awschalom, J. J. Berry, and N. Samarth,
Phys. Rev. Lett. {\bf 84}, 1015 (2000).
\bibitem{RS91IP97} Lowering the symmetry results in new terms in the spin-orbit coupling Hamiltonians and, hence, in the new mechanisms of spin-orbit coupling. For review see E. I. Rashba and V. I. Sheka, in: {\it Landau Level Spectroscopy}, G. Landwehr and E. I. Rashba eds. (North-Holland, Amsterdam, 1991) p. 131; E. L. Ivchenko and G. E. Pikus, {Superlattices and Other Heterostructures}, 2nd ed. (Springer, New York, 1997).
\bibitem{KVOles} O. Krebs and P. Voisin, Phys. Rev. Lett. {\bf 77}, 1829 (1996); J. T. Olesberg, W. H. Lau, M. E. Flatt\'{e}, C. Yu, E. Altunkaya, E. M. Shaw, T. C. Hasenberg, and T. F. Boggess, Phys. Rev. B {\bf 64}, 201301 (2001);  R. M. Stroud, A. T. Hanbicki, Y. D. Park, A. G. Petukhov, B. T. Jonker, G. Itskos, G. Kioseoglou, M. Furis, A. Petrou, cond-mat/0110570.
\bibitem{MPT76} R. Meservey, D. Paraskevopoulos, and P. M. Tedrow, Phys. Rev. Lett. {\bf 37}, 858 (1976).
\bibitem{WG00} D. C. Worledge and T. H. Geballe, Phys. Rev. Lett. {\bf 85}, 5182 (2000).
\bibitem{HA73} J. A. Hertz and K. Aoi, Phys. Rev. B {\bf 8}, 3252 (1973).
\bibitem{Teresa99} J. M. De Teresa, A. Barth\'{e}l\'{e}my, A. Fert, J. P. Contour, R. Lyonnet, F. Montaigne, P. Seneor, and A. Vaur\`{e}s, Phys. Rev. Lett. {\bf 82}, 4288 (1999).
\bibitem{M01} I. I. Mazin, Europhys. Lett. {\bf 55}, 404 (2001).
\bibitem{Ploog02} K. H. Pooog, J. Appl. Phys. {\bf 91}, 7256 (2002). 
\bibitem{ELS02} S. C. Ervin, S.-H. Lee, and M. Scheffer, Phys. Rev. B {\bf 65}, 205422 (2002).
\bibitem{Bloch} P. Mavropoulos, N. Papanikolaou, and P. H. Dederichs,
Phys. Rev. Lett. {\bf 85}, 1088 (2000); M. Zwierzycki, K. Xia, P. J. Kelly, 
G. E. W. Bauer, and I. Turek, cond-mat/0204422. 
\bibitem{G01} D. Grundler, Phys. Rev. B {\bf 63}, R161307 (2001).
\bibitem{SM02} G. Schmidt and L. W. Molenkamp, Semicond. Sci. Technol. {\bf 17}, 310 (2002).
\bibitem{HSNT01} H. B. Heersche, Th. Sch\"{a}pers, J. Nitta, and H. Takayanagi,  Phys. Rev. B {\bf 64}, 161307 (2001).
\bibitem{KR02} V. Ya. Kravchenko and E. I. Rashba, cond-mat/0209539.
\end{references}
\end{document}